\documentclass[12pt,a4paper]{article}
\pdfoutput=1

\usepackage[utf8]{inputenc}
\usepackage[T1]{fontenc}
\usepackage[margin=2.5cm]{geometry}
\usepackage[pdfstartview={FitH},bookmarks=false,linktoc=page,colorlinks=true,linkcolor=blue,citecolor=blue,urlcolor=blue]{hyperref}
\usepackage[numbered]{bookmark}
\usepackage{mathtools}
\usepackage{amssymb}
\usepackage{graphicx}
\usepackage[font=small,labelfont=bf]{caption}
\usepackage{authblk}
\usepackage{cite}
\usepackage{dsfont}
\usepackage{xcolor}
\usepackage[titles]{tocloft}
\usepackage{float}
\usepackage{subcaption}
\usepackage{color}
\usepackage{tikz}
\usepackage{ifthen}
\usepackage[warn]{textcomp}
\numberwithin{equation}{section}
\allowdisplaybreaks

\begin{document}

\title{\vspace{2cm}\textbf{Shadows of Black Holes in Vector-Tensor Galileons Modified Gravity}\vspace{1cm}}

\author[a]{Tsvetan Vetsov}
\author[a]{Galin Gyulchev}
\author[a,b]{Stoytcho Yazadjiev}
\affil[a]{\textit{Department of Physics, Sofia University,}\authorcr\textit{5 J. Bourchier Blvd., 1164 Sofia, Bulgaria}}

\affil[b]{\textit{Theoretical Astrophysics, Eberhard Karls University of T\"ubingen,}\authorcr\textit{T\"ubingen 72076, Germany}

\vspace{10pt}\texttt{vetsov,gyulchev,yazad@phys.uni-sofia.bg}\vspace{0.1cm}}
\date{}
\maketitle

\begin{abstract}
We study the shadows of disformal black holes in vector-tensor Galileons modified gravity. Our analysis shows that the apparent image of the black hole in the observer's sky is non-spherical and cuspy, which is in contrast to the Kerr and Kerr-Newman cases. The non-trivial silhouette of the apparent image of the black hole provides novel templates for the current astrophysical observations. Moreover, due to the non-minimal coupling of the vector field to gravity the disformal black hole supports regular horizons for spin parameter exceeding the ADM mass. Finally, the shadows of the  massless limit, supported only by the charge of the dark vector field, are also studied. When relevant the results are compared with the Kerr black hole.
\end{abstract}

\vspace{0.5cm}
\vspace{0.5cm}
\thispagestyle{empty}

\noindent\rule{\linewidth}{0.75pt}
\vspace{-0.8cm}\tableofcontents
\noindent\rule{\linewidth}{0.75pt}

\section{Introduction}\label{sec: Introduction}

Black holes always present great challenges in modern physics. Studying the properties of space-time and matter near the event horizon of a black hole is highly non-trivial task theoretically and experimentally. From theoretical point of view, the existence of many physically viable models of gravity with predictions deviating from Einstein's General Theory of Relativity (GR) poses the natural question which theory is realized in nature. On the other hand, the extremely small size of the black holes and their relatively large distance from us make them very hard to observe and study.

However, the hope is that the advancement of nowadays astrophysical and gravitational wave observations will shed light into previously inaccessible properties of these cosmic phenomena. Most promising experiments capable of such observations include the Laser Interferometer Gravitational-Wave Observatory (LIGO) \cite{Abbott:2016blz, PhysRevLett.116.241103, PhysRevLett.118.221101, PhysRevLett.119.141101, Abbott:2017gyy}, from gravitational waves perspective, and the Event Horizon Telescope (EHT) \cite{Doeleman:2008qh}, from electromagnetic perspective. It is expected that the data from EHT  mission will be precise enough to measure some intrinsic properties of the supermassive compact object at the center of our galaxy. Most interesting observational feature of the black hole is the shadow it casts when illuminated by some nearby sources of light. These observations will outline important properties of space-time in strong gravitational regime and put to test any deviations from the standard Kerr geometry. The first real images of a black hole shadow are expected to arrive very soon. Therefore it is important to conduct thorough analytic and numerical investigations on the apparent shape of different black hole configurations, thus providing new shadow templates for the ongoing observations.

Many such examples already exist in the literature including the Kerr-Newmann family of solutions of the Einstein-Maxwell equations \cite{Bardeen:1973tla, chandrasekhar1992mathematical, Takahashi:2004xh, PhysRevD.80.024042}; the shadow of a black
hole with NUT-charges \cite{Chakraborty:2013kza, Grenzebach:2014fha}; the black hole shadows in Einstein-Maxwell-dilaton gravity \cite{PhysRevD.87.044057, Wei:2013kza}; Chern-Simons modified gravity \cite{PhysRevD.81.124045}; braneworld gravity \cite{PhysRevD.85.064019, Eiroa:2017uuq}; the apparent shape of the Sen black hole \cite{PhysRevD.78.044007, Dastan:2016bfy, Younsi:2016azx}. More interesting examples of black hole shadows also include colliding and multi- black holes\cite{Nitta:2011in, Yumoto:2012kz}; rotating black holes in $f(R)$ gravity \cite{Dastan:2016vhb}; conformal Weyl gravity \cite{Mureika:2016efo}; Einstein-dilaton-Gauss-Bonnet black holes \cite{Cunha:2016wzk}; higher-dimensional black holes \cite{Papnoi:2014aaa, Singh:2017vfr, Amir:2017slq}; non-commutative geometry
inspired black holes \cite{Wei:2015dua, Sharif:2016znp}; Einstein-Born-Infeld black
holes \cite{Atamurotov:2015xfa};  Ayon-Beato-Garcia black hole \cite{Abdujabbarov:2016hnw}; rotating
Hayward and rotating Bardeen regular black holes \cite{Abdujabbarov:2016hnw}; hairy black holes \cite{Cunha:2015yba, Cunha:2016bjh}; chaotic shadow of a non-Kerr rotating compact objects with quadrupole mass moment and a magnetic dipole \cite{Wang:2018eui, Wang:2017qhh}; and black holes with exotic matter \cite{Tinchev:2015apf, Abdujabbarov:2015pqp, Singh:2017xle, Huang:2016qnl}. Shadows of wormholes and naked singularities have also been investigated in \cite{Nedkova:2013msa, Ohgami:2015nra, Ortiz:2015rma}.

Recently an exact analytic rotating black hole solution with regular horizons was obtained by F. Filippini and G. Tasinato in \cite{TS:2017}. It is a solution in type vector-tensor Galileons modified gravity \cite{Heisenberg:2014rta, Jimenez:2016isa, Heisenberg:2017xda, Heisenberg:2017hwb} derived by a disformal transformation \cite{Bekenstein:1992pj, Bettoni:2013diz, Zumalacarregui:2013pma, Kimura:2016rzw} on a version of the Kerr-Newman solution of the Einstein-Maxwell theory of gravity. The model is a modification of Einstein gravity with additional vector degrees of freedom, which can be associated
with dark matter or dark energy. Our main goal will be to investigate
the apparent shape of this new rotating black hole solution for specific subsets in its parameter space.

This paper is organized as follows. In Section \ref{sec:2} we shortly discuss the disformal black hole solution \cite{TS:2017}, its symmetries and its horizons. In Section \ref{sec:3} we choose the basis for the local observer and parametrize the observer's plane. We also discuss some common features of the black hole shadow. In Section \ref{sec:4} we present our study on the shadow silhouette cast by the rotating disformal solution derived in \cite{TS:2017}. Finally, in Section \ref{sec:conclusion} we make a short summary of our results.

\section{Rotating disformal black hole solution}\label{sec:2}

\subsection{General setup}

The starting point is the Einstein-Maxwell type of action given by
\begin{equation}\label{eq:Einstein-Maxwell_Action}
{S_{EM}} = \int {d^4}x\,\sqrt { - g} \,\left( {\frac{R}{4} - \frac{1}{4}\,{F_{\mu \nu }}\,{F^{\mu \nu }}} \right)\,.
\end{equation}
One well-known solution to (\ref{eq:Einstein-Maxwell_Action}) is the Kerr-Newman solution in Boyer-Lindquist coordinates $(t, r, \theta, \varphi)$:
\begin{equation}\label{eq:Kerr_Newman_solution}
d{s^2} = \left( {\frac{{d{r^2}}}{\Delta } + d{\theta ^2}} \right)\,\Sigma  - {(dt - a\,{\sin ^2}\theta \,d\varphi )^2}\,\frac{\Delta }{\Sigma } + {\left( {({r^2} + {a^2})\,d\varphi  - a\,dt} \right)^2}\,\frac{{{{\sin }^2}\theta }}{\Sigma }
\end{equation}
with gauge field given by
\begin{equation}\label{eq:Einstein-Maxwell_gauge_field}
{A_\mu } = \left( { - \frac{{Q\,r}}{\Sigma },\,0,0,\frac{{a\,Q\,r\,{{\sin }^2}\theta }}{\Sigma }} \right)\,.
\end{equation}
To generate new solutions one can act on KN solution (\ref{eq:Kerr_Newman_solution}) with a disformal transformation \cite{Bekenstein:1992pj, Bettoni:2013diz, Zumalacarregui:2013pma, Kimura:2016rzw} involving new vector fields and parametrized by a real constant $\beta$, namely
\begin{equation}\label{eq:Disformal_transformation}
{{\tilde g}_{\mu \nu }}(x) = {g_{\mu \nu }}(x) - {\beta ^2}\,{A_\mu }(x)\,{A_\nu }(x)\,.
\end{equation}
Here, the components of the vector field are considered to be gauge fields of some dark force and should not be associated with the standard Maxwell $U(1)$ gauge field. The configuration obtained in \cite{TS:2017} excites the radial component $A(r)$ of the gauge field (\ref{eq:Einstein-Maxwell_gauge_field}) and after a suitable ansatz, namely $A(r)=Q\,r/\Delta(r)$, results in the following rotating black hole solution
\begin{align}\label{eq:TS_solution}
\nonumber d{s^2} &= \left( {\frac{\Sigma }{{\Delta \,\Sigma  - {\beta ^2}\,{Q^2}\,{r^2}}}\,d{r^2} + d{\theta ^2}} \right)\,\Sigma  - {(dt - a\,{\sin ^2}\theta \,d\varphi )^2}\,\frac{{\Delta \,\Sigma  + {\beta ^2}\,{Q^2}\,{r^2}}}{{{\Sigma ^2}}}
\\
&+ {\left( {({r^2} + {a^2})\,d\varphi  - a\,dt} \right)^2}\,\frac{{{{\sin }^2}\theta }}{\Sigma }\,,
\end{align}
which is asymptotically flat and charged under the dark vector field\footnote{The vector field profile (\ref{eq:dark_gauge_field}) has three physical components turned on, against the two of the Kerr-
Newman configuration. The vector radial component is physical in this case and can not be gauged
away without simultaneously changing the metric.}
\begin{equation}\label{eq:dark_gauge_field}
{A_\mu } = \left( { - \frac{{Q\,r}}{\Sigma },\,\frac{{Q\,r\,\Sigma }}{{\Delta \,\Sigma  - {\beta ^2}\,{Q^2}\,{r^2}}},0,\frac{{a\,Q\,r\,{{\sin }^2}\theta }}{\Sigma }} \right)\,.
\end{equation}
One also has the $\Delta$ and $\Sigma$ functions given by
\begin{equation}
\Delta  = {a^2} + {r^2} - 2\,M\,r + {Q^2},\qquad \Sigma  = {r^2} + {a^2}\,{\cos ^2}\theta \,.
\end{equation}
The new solution (\ref{eq:TS_solution}) satisfies the equations of motions of specific vector-tensor Galileons type of theory with the following disformed action (up to total derivative):
\begin{align}\label{eq:Disformal_action}
\nonumber
{S_{disf}} &= \int {{d^4}x} \,\frac{{\sqrt { - g} }}{{4\,{\gamma _0}}}\,\left( {R - \frac{{{\beta ^2}}}{4}\,\gamma _0^2\,({S_{\mu \nu }}\,{S^{\mu \nu }} - {S^2}) - \frac{{4 - {\beta ^2}}}{4}\,{F_{\mu \nu }}\,{F^{\mu \nu }}} \right.
\\
&\left. { + \frac{{{\beta ^2}\,({\beta ^2} - 4)}}{2}\,\gamma _0^2\,{F_{\mu \rho }}\,F_\nu ^\rho \,{A^\mu }\,{A^\nu }} \right)\,,
\end{align}
where one has defined the following notations:
\begin{equation}
{F_{\mu \nu }} = {\nabla _\mu }{A_\nu } - {\nabla _\nu }{A_\mu }\,,\quad {S_{\mu \nu }} = {\nabla _\mu }{A_\nu } + {\nabla _\nu }{A_\mu },\quad S = {S_{\mu \nu }}\,{g^{\mu \nu }}\,.
\end{equation}
It contains non-minimal couplings of the
vector field to gravity with strength controlled by the values of the parameter $\beta$. Here, the disformal coupling $\beta$ induces deviations from the Kerr-Newman geometry and also affects the apparent image of the black hole. The action (\ref{eq:Disformal_action}) also contains derivative self-interactions of the form that usually appears in Horndeski systems.

\subsection{Horizons and symmetries}

\begin{figure}[H]
    \centering
    \begin{subfigure}[b]{0.45\textwidth}
        \includegraphics[width=\textwidth]{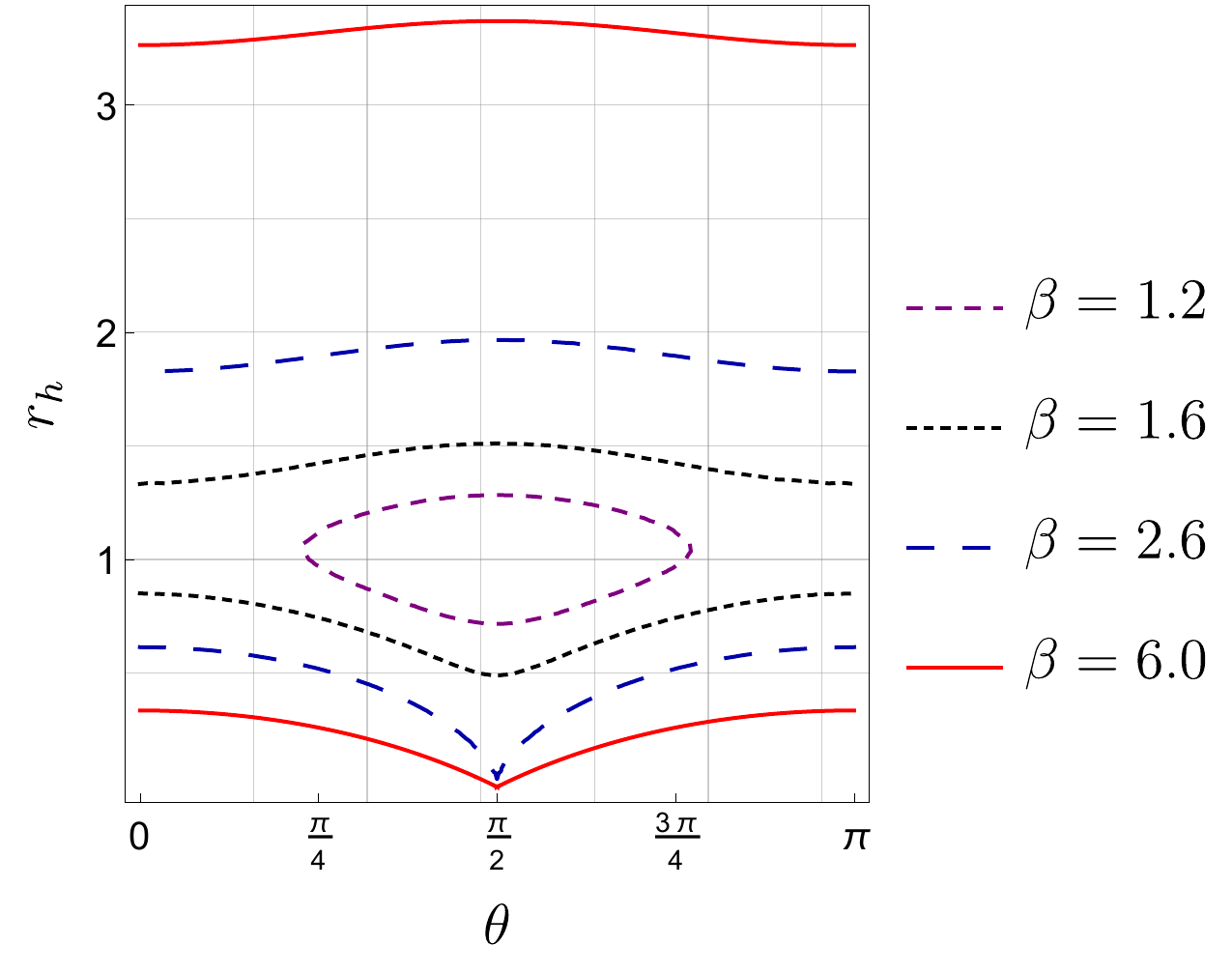}
        \caption{ $a=0.995$.}
        \label{fig:TS_rth1a}
    \end{subfigure}
\hspace{0.8 mm}
    \begin{subfigure}[b]{0.45\textwidth}
        \includegraphics[width=\textwidth]{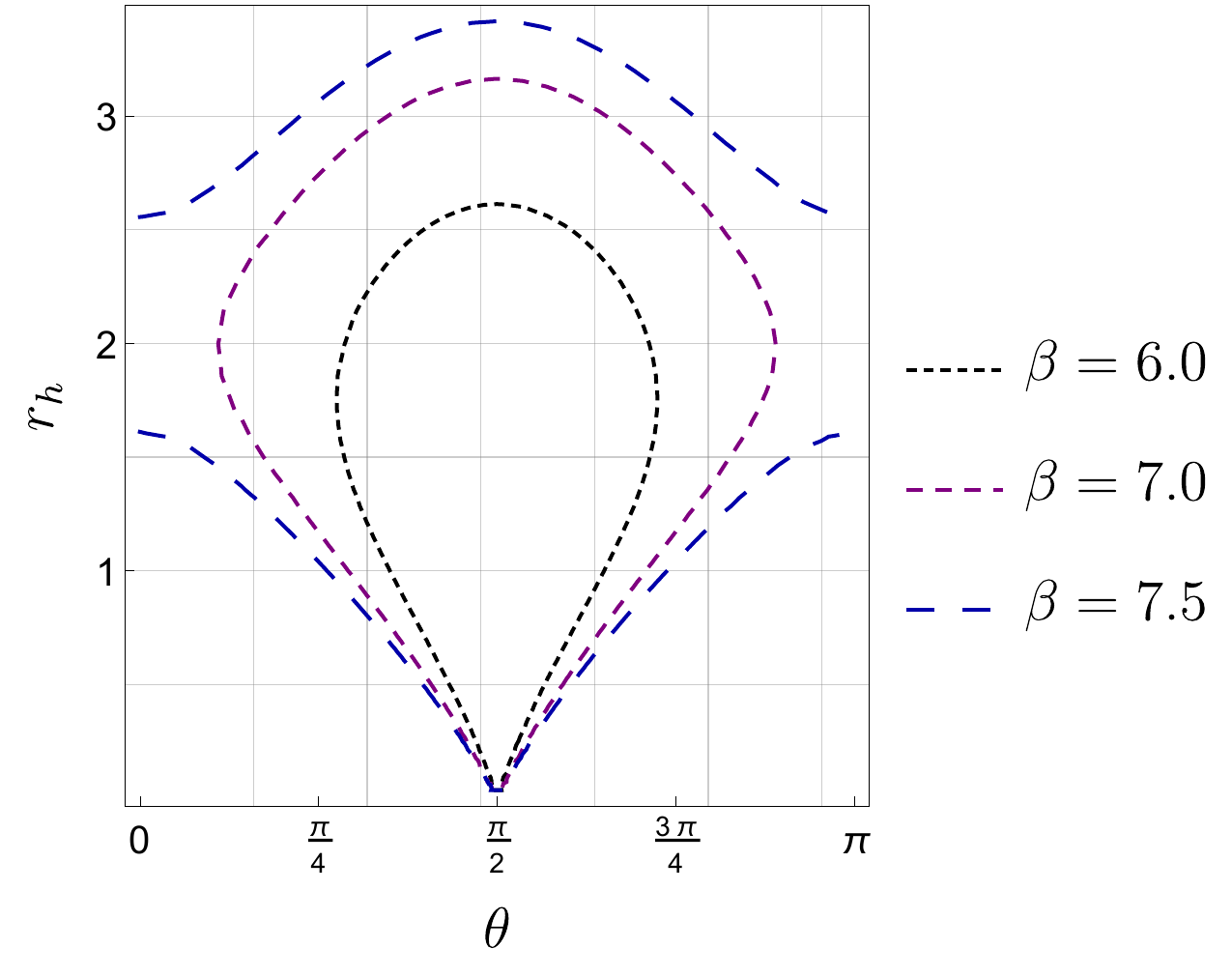}
        \caption{ $a=2.0$.}
        \label{fig:TS_rth1b}
    \end{subfigure}
    \caption{The dependence of the Cauchy horizon (lower curves) and the event horizon (upper curves) on the angle $\theta$ for $M=1$, $Q=0.4$ and different black hole spins.}\label{fig:TS_rth1}
\end{figure}

\begin{figure}[H]
    \centering
    \begin{subfigure}[b]{0.226\textwidth}
        \includegraphics[width=\textwidth]{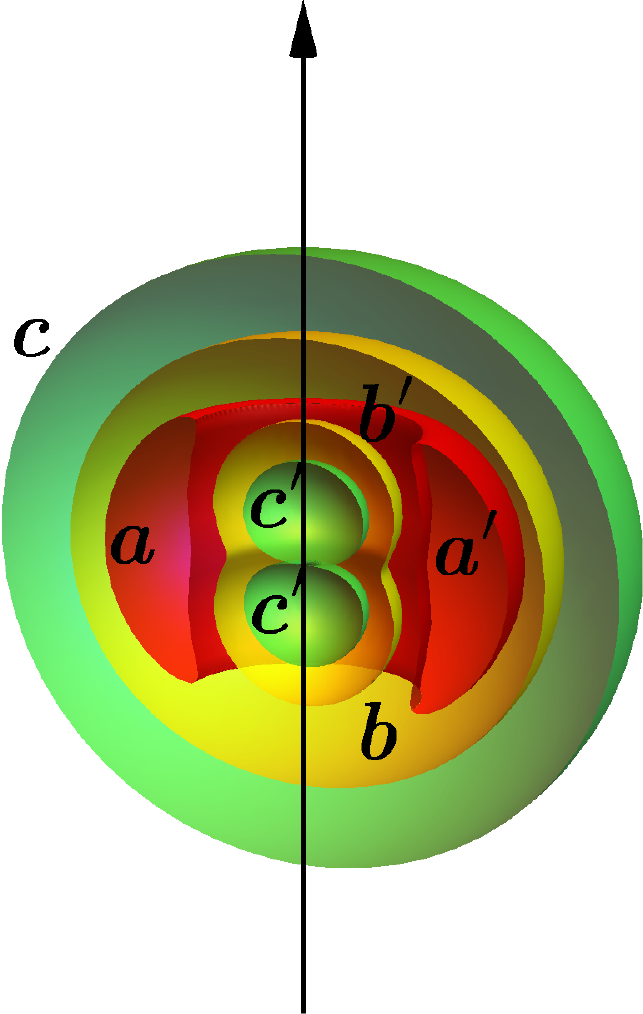}
        \caption{ $a=0.995$.}
        \label{fig:}
    \end{subfigure}
\hspace{1.75cm}
    \begin{subfigure}[b]{0.4\textwidth}
        \includegraphics[width=\textwidth]{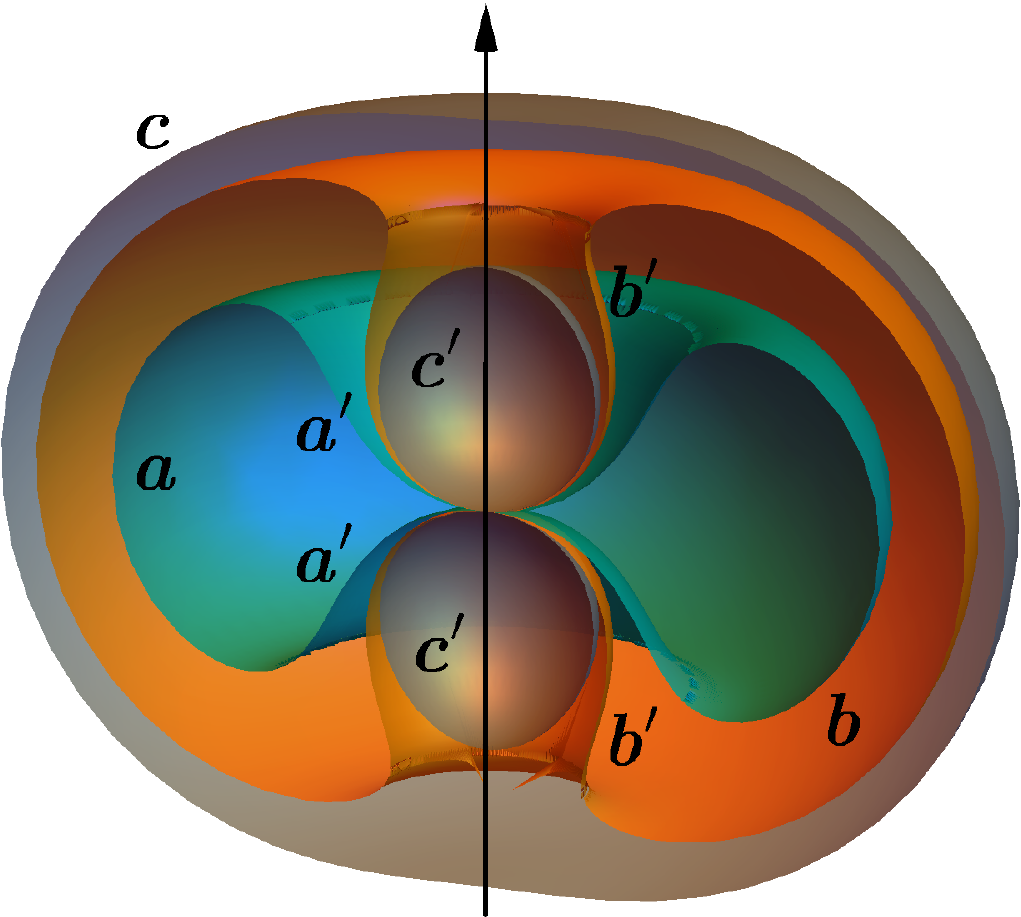}
        \caption{ $a=2.0$.}
        \label{fig:}
    \end{subfigure}
    \caption{An actual 3D model of the horizons of the disformal black hole sliced vertically through the poles. Figure {\bf{(a)}} on the left shows the surfaces, defining the inner Cauchy horizon and the outer event horizon, for values of $\beta$ corresponding to the level curves, shown in Fig. \ref{fig:TS_rth1a}. Here, the $a'$ and $a$ surfaces (red) correspond to the inner Cauchy horizon and the event horizon respectively, for $\beta=1.2$. The $b'$ and $b$ surfaces (yellow) correspond to the inner and the outer horizons for $\beta=1.6$. The surfaces, labeled as $c'$ and $c$ (green), depict the $\beta=2.6$ case. Figure {\bf{(b)}} on the right shows the horizons for the values of $\beta$ in the ultraspining case from Fig. \ref{fig:TS_rth1b}. Here, the $a'$ and $a$ surfaces (blue) correspond to the inner  and the outer horizon for $\beta=6.0$. The $b'$ and $b$ surfaces (orange) are for $\beta=7.0$. Finally, the surfaces, labeled as $c'$ and $c$ (grey), correspond to the horizons for $\beta=7.5$. From the illustrations above one can notice the oblate shape of the event horizon. For large values of the disformal parameter $\beta$ the shape of the event horizon approaches spherical one in both cases. }\label{fig:TS_rth3}
\end{figure}

\begin{figure}[H]
    \centering
    \begin{subfigure}[b]{0.45\textwidth}
        \includegraphics[width=\textwidth]{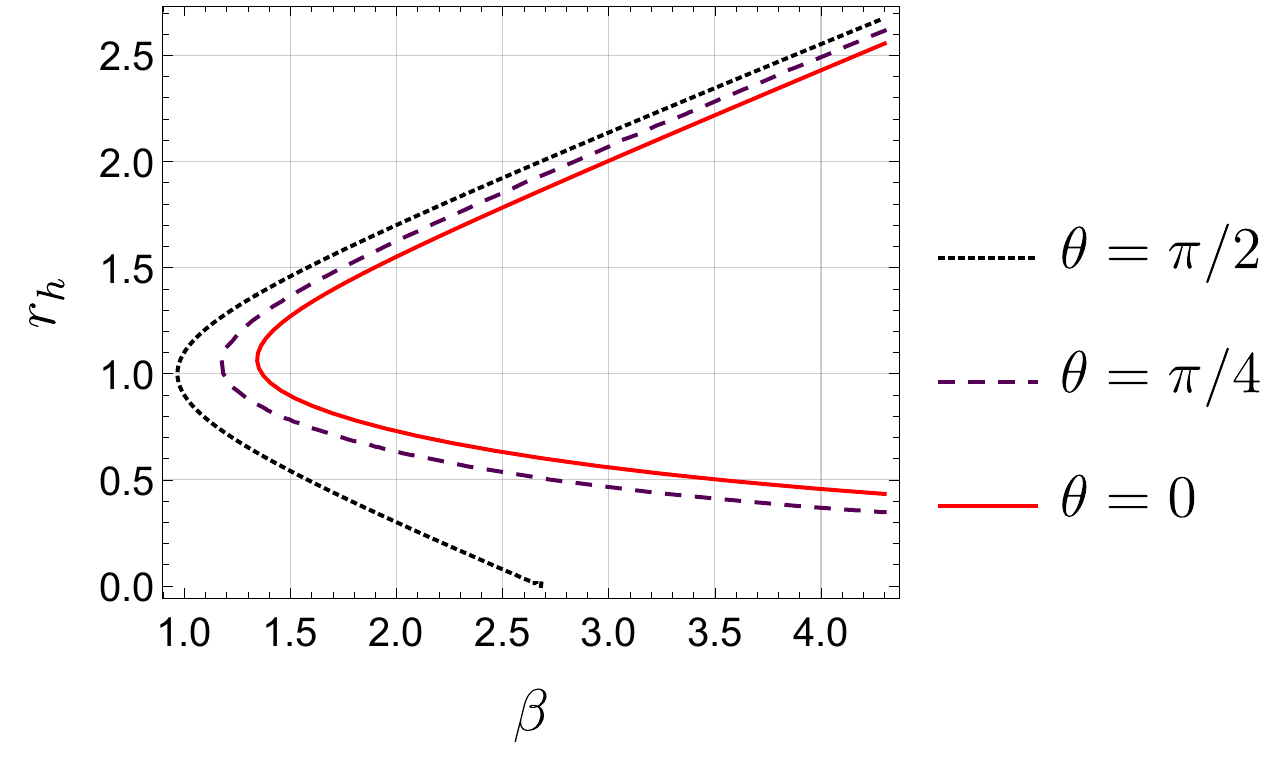}
        \caption{ $a=0.995$.}
        \label{fig:}
    \end{subfigure}
\hspace{0.8 mm}
    \begin{subfigure}[b]{0.45\textwidth}
        \includegraphics[width=\textwidth]{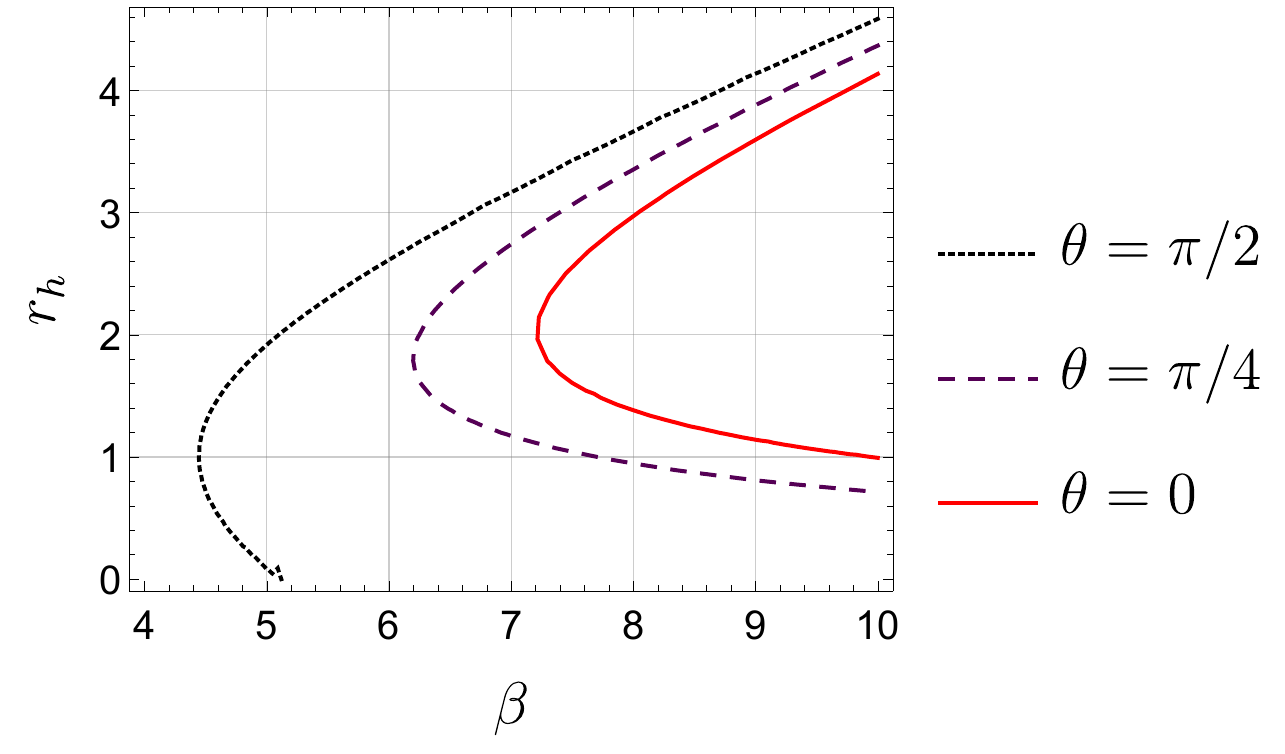}
        \caption{ $a=2.0$.}
        \label{fig:}
    \end{subfigure}
    \caption{The dependence of the Cauchy horizon (lower curves) and the event horizon (upper curves) on the disformal parameter $\beta$ for $M=1$, $Q=0.4$ and different black hole spins.}\label{fig:TS_rth2}
\end{figure}

The event horizon is located at radial distance given by the roots of $\partial^\mu r\,\partial_\mu r=g^{rr}=0$:
\begin{equation}\label{HorizonEq}
{r^4} - 2\,M\,{r^3} + \left( {{a^2}\,(1 + {{\cos }^2}{\theta}) - q\,{M^2}} \right){r^2} - 2\,{a^2}\,M\,{\cos ^2}{\theta}\,r + {a^2}\,{\cos ^2}{\theta}\,\left( {{a^2} + \frac{{q\,{Q^2}\,{M^2}}}{{1 - {\beta ^2}}}} \right) = 0\,,
\end{equation}
where the effective parameter
\begin{equation}
q = \frac{{{Q^2}}}{{{M^2}}}\,({\beta ^2} - 1)
\end{equation}
can be positive or negative, depending on the size of $\beta$. This is a fourth order in the coordinate $r$ algebraic equations. Depending on the sign of the discriminant, it can have
four, two or no real roots. One can also discern two regimes when $\beta<1$ and $\beta>1$, the latter corresponding to strong non-minimal coupling between the vector field and gravity. In the strong coupling regime, $\beta>1$, the black hole spin parameter $a$ can exceed $M$ and thus solution (\ref{eq:TS_solution}) is valid for arbitrary values of $a$. In this ultraspinning case the disformed black hole can still maintain its horizons due to the non-minimal vector-tensor interactions, which are able to contrast strong centrifugal forces. The price one has to pay translates into deformations of the horizon shape, which seems to also have an imprint on the contour of the shadow as shown in Section \ref{sec:4}. On Figures \ref{fig:TS_rth1} and \ref{fig:TS_rth2} are portrayed some level curves in the ($r,\theta$) plane of the equation $g^{rr}=0$ for $a<M$ and $a>M$. We have also constructed an actual 3D model of the horizons for the disformal black hole in Boyer-Lindquist coordinates as depicted on Fig. \ref{fig:TS_rth3}. The shape of the horizons follow the behaviour of the level curves from Fig. \ref{fig:TS_rth1}.
One notices that the radial position of the external event horizon $r_h$ depends on $\theta$, thus there is a difference in the radius of the horizon at the poles and at the equator. The ratio between the radial size of the horizon at the poles versus the size of the horizon at the equator quantifies the oblateness $\omega$ of the black hole:
\begin{equation}
\omega  = 1 - \frac{{r_h^{pol}}}{{r_h^{eq}}}\,.
\end{equation}

The fact that the components of the metric does not depend on $t$ and $\varphi$ leads to the existence of one time-like and one space-like Killing vector, namely $K_{t}=\partial_t$ and $K_{\varphi}=\partial_\varphi$. This implies two conserved charges, namely the energy, $p_t=-E$, and the angular momentum, $p_\varphi=L$, of the system. The hyper-surface, corresponding to $K_{(t)\mu} \,K_{(t)}^{\mu}=0$, defines the ergosphere. The equation for the ergosphere is also fourth order in $r$ algebraic equation. As pointed out in \cite{TS:2017} the ergosphere always lies outside the outer event horizon.

\section{Observers and impact parameters}\label{sec:3}

\subsection{Local observer}

For an observer, the black hole shadow is its apparent image in the sky due to the gravitational lensing of nearby radiation emitted by some external light sources \cite{Bardeen:1973tla}. We choose to work in the observer basis given in \cite{2016IJMPD}.

Let ($t,\,r,\,\theta,\,\varphi$) are the Boyer-Lindquist spherical-like coordinates. One can expand the local observer basis $(\hat e_{(t)},\,\hat e_{(r)},\,\hat e_{(\theta)},\,\hat e_{(\varphi)})$ in the coordinate vector basis ($\partial_t,\,\partial_r,\,\partial_\theta,\,\partial_\varphi$) in the following way:
\begin{equation}\label{eq:Local_observer_basis_decomposition}
\hat e_{(t)}=\zeta\,\partial_t+\gamma\,\partial_\varphi\,,\qquad \hat e_{(r)}=A^r\,\partial_r\,,\qquad \hat e_{(\theta)}=A^\theta\,\partial_\theta\,,\qquad \hat e_{(\varphi)}=A^\varphi\,\partial_\varphi\,.
\end{equation}
Note that the chosen decomposition is not unique, allowing for spatial rotations and Lorentz boosts. This particular choice is connected to the ZAMO (zero angular momentum observers) reference frame \cite{Frolov:1998}. Using Minkowski normalization, $\hat e_{(\mu)}\cdot\hat e_{(\nu)}=\eta_{\mu\nu}$, one can find the coefficients in the decomposition (\ref{eq:Local_observer_basis_decomposition}):
\begin{equation}
A^\theta=\frac{1}{\sqrt {g_{\theta\theta}}}\,,
\qquad A^r=\frac{1}{\sqrt {g_{rr}}}\,,\qquad
A^\varphi=\frac{1}{\sqrt {g_{\varphi\varphi}}}\,,
\end{equation}
and
\begin{equation}
\gamma=-\frac{g_{t\varphi}}{g_{\varphi\varphi}}\,\sqrt{\frac{g_{\varphi\varphi}}{g^2_{t\varphi}-g_{tt}\,g_{\varphi\varphi}}}\,,\qquad
\zeta=\sqrt{\frac{g_{\varphi\varphi}}{g^2_{t\varphi}-g_{tt}\,g_{\varphi\varphi}}}\,.
\end{equation}
The locally measured momenta of the photon can also be obtained:
\begin{equation}
p^{(t)}=-\hat e^{\mu}_{(t)}\,p_\mu=E\,\zeta-L\,\gamma\,,\qquad p^{(r)}=\hat e^{\mu}_{(r)}\,p_\mu=\frac{p_r}{\sqrt{g_{rr}}}\,,\quad
\end{equation}
and
\begin{equation}
p^{(\theta)}=\hat e^{\mu}_{(\theta)}\,p_\mu=\frac{p_\theta}{\sqrt{g_{\theta\theta}}}\,,\qquad
p^{(\varphi)}=\hat e^{\mu}_{(\varphi)}\,p_\mu=\frac{L}{\sqrt{g_{\theta\theta}}}\,.
\end{equation}

\subsection{Impact parameters}

The projection of photons detected in an image plane corresponds to the optical perspective of an observer. The Cartesian coordinates $(x, y)$ assigned to each photon in this image plane are its impact parameters \cite{0004-637X-773-1-57} and they are proportional to the respective observation angles $(\hat\alpha,\,\hat \beta)$ \cite{2016IJMPD}:
\begin{equation}
x\equiv -\tilde r \,\hat\beta\,,\qquad y\equiv\tilde r\,\hat\alpha\,,
\end{equation}
where the perimetral radius $\tilde r$ is  defined as $\tilde r=\mathcal{P}/(2\,\pi)=\sqrt{g_{\varphi\varphi}}$ and computed at the position of the observer. The angular coordinates $(\hat\alpha,\,\hat \beta)$ of a point in the observer's plane define the direction of the associated light ray and
establishes its initial conditions. The photon momenta can also be parametrized in terms of the observable angles $(\hat\alpha,\,\hat \beta)$ such as
\begin{equation}
p_\theta=\sqrt{g_{\theta\theta}}\, \sin\hat\alpha\,,\qquad L=\sqrt{g_{\varphi\varphi}}\, \sin\hat\beta\,\cos\hat\alpha\,,
\end{equation}
and
\begin{equation}
p_r=\sqrt{g_{rr}}\, \cos\hat\beta\,\cos\hat\alpha\,,\qquad E=\frac{1+\gamma \,\sqrt{g_{\varphi\varphi}}\, \sin\hat\beta\,\cos\hat\alpha}{\zeta}\,.
\end{equation}
Having established the observer and the observer's plane one can now proceed with the numerical integration of the Hamilton equations for the null geodesics in the considered rotating background (\ref{eq:TS_solution}), namely
\begin{equation}
\dot x^\mu=\frac{\partial H}{\partial p_\mu}\,,\qquad \dot p_\mu=\frac{\partial H}{\partial x^\mu}\,,
\end{equation}
where the Hamiltonian and the conserved momenta are given by
\begin{equation}
H=\frac{1}{2}\,g^{\mu\nu}\, p_\mu\,p_\nu=0\,,\qquad p_t=-E\,,\qquad p_\varphi=L\,.
\end{equation}
The resulting shadow contours of the disformal black hole are shown in Section \ref{sec:4}.

\subsection{Common characteristics of the black hole shadow}

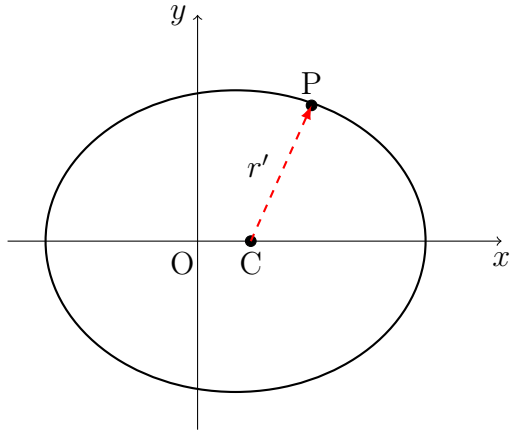
\begin{figure}[H]
\begin{center}
\begin{tikzpicture}
    \draw[thin,->,black] (-2.5,0)--(4,0) node[below] {$x$}; 
    \draw[thin,->,black] (0,-2.5)--(0,3) node[left] {$y$}; 
    \draw[black,thick] (0.5,0) ellipse (2.5cm and 2cm);

\coordinate (C) at (0.7,0);
\coordinate (P) at (1.5,1.8);
\coordinate (O) at (0,0);

\draw [fill=black] (C) circle (2pt) node [below] {C};
\draw [fill=black] (P) circle (2pt) node [above] {P};

\draw [-latex, red, thick, dashed] (C) -- (P);

\draw node[below,xshift=-0.2cm] {O};

 \node[] at (0.8, 1) (b) {$r'$};
\end{tikzpicture}
\caption{A graphical depiction of the contour of a black hole shadow in the $(x, y)$ image plane of the observer.\label{fig:SBH}}
\end{center}
\end{figure}

Here we introduce some useful geometric features of a generic black hole shadow (Fig. \ref{fig:SBH}).
First of all, the center of the black hole is given by the point $C$, as shown on Figure {\ref{fig:SBH}}. Its abscissa is determined by $x_C=(x_{max}+x_{min})/2$, where $x_{min}$ and $x_{max}$ are respectively the minimum and
the maximum abscissa of the shadow’s edge. Since the points $C$ and $O$ generally do not coincide, $x_C$ can be considered as a specific feature of the shadow.
Secondly, any generic point $P$ on the shadow’s contour is at a distance $r'$ from $C$, which is defined as the Euclidian distance $r'=\sqrt{y_P^2+(x_P-x_C)^2}$ on the observer's plane. Given the line element $ds^2=dx^2+dy^2$, one can define further useful geometric features of the apparent image, namely the perimeter $\mathcal{P}$ of the
shadow, its average radius $\bar r_{sh}$ and the deviation from sphericity $\sigma_r$ \cite{Cunha:2016wzk, Abdujabbarov:2015xqa, grenzebach2016shadow}:
\begin{equation}
\mathcal{P}\equiv\oint ds\,,\qquad \bar r_{sh}\equiv\frac{1}{\mathcal{P}}\,\oint{r'\,ds}\,,\qquad \sigma_r=\left(\frac{1}{\mathcal{P}}\,\oint\left(1-\frac{r'}{\bar r_{sh}}\right)^2\,ds\right)^{1/2}\,.
\end{equation}
All these parameters are expressed in units of the ADM mass $M$. In some cases, it is possible to compare the shadow parameters of the disformal solution (\ref{eq:TS_solution}) with those from the Kerr black hole with the same mass $M$ and spin $a$.
Hence, one can also define the relative deviations to the Kerr case in the following way:
\begin{equation}
\delta_{\bar r_{sh}}=\frac{\bar r_{sh}-\bar r_{Kerr}}{\bar r_{Kerr}}\,,\qquad \delta_{\sigma_{r}}=\frac{\sigma_r-\sigma_{Kerr}}{\sigma_{Kerr}}\,,\qquad \delta_{x_C}=\frac{x_C-x_{C\,Kerr}}{x_{C\,Kerr}}\,.
\end{equation}
In Tables \ref{table:one} and \ref{table:two} are shown specific values, which capture the general features of the considered disformal black hole shadow.

In what follows we will numerically analyze the shadow of the black hole given by Eq. (\ref{eq:TS_solution}) for different subsets of the parameter space ($M, Q, a, \beta$). The existence of two conserved quantities facilitate the numerical calculations, which are conducted on the Wolfram Mathematica computer algebra system.

\section{Black hole shadows of the disformal solution}\label{sec:4}

The form of the solution (\ref{eq:TS_solution}) does not allow for separation of variables in the Hamilton-Jacobi equation for the null geodesics. Therefore one is forced to study the contour of the black hole shadow numerically. Our analysis shows that the apparent image of the Filippini-Tasinato disformal black hole solution (\ref{eq:TS_solution}) is non-spherical and cuspy. Such cuspy
silhouette of the shadow emerges in the space-time of hairy black hole \cite{PhysRevD.96.024039} and
rotating non-Kerr black hole \cite{1475-7516-2017-10-051}. The recent investigation indicates that these novel
structure and patterns in the shadows are determined actually by the non-planar bound photon orbits \cite{PhysRevD.96.024039} and
the invariant phase space structures \cite{PhysRevD.96.024045} for the photon motion in the given background space-times. These features depict a major qualitative difference with respect to Kerr and Kerr-Newman solutions and give potentially new templates for the current observations of black holes and other compact objects.

\subsection{Shadows for $0<\beta<1$ (weak non-minimal coupling regime)}

\begin{figure}[H]
    \centering
    \begin{subfigure}[b]{0.45\textwidth}
        \includegraphics[width=\textwidth]{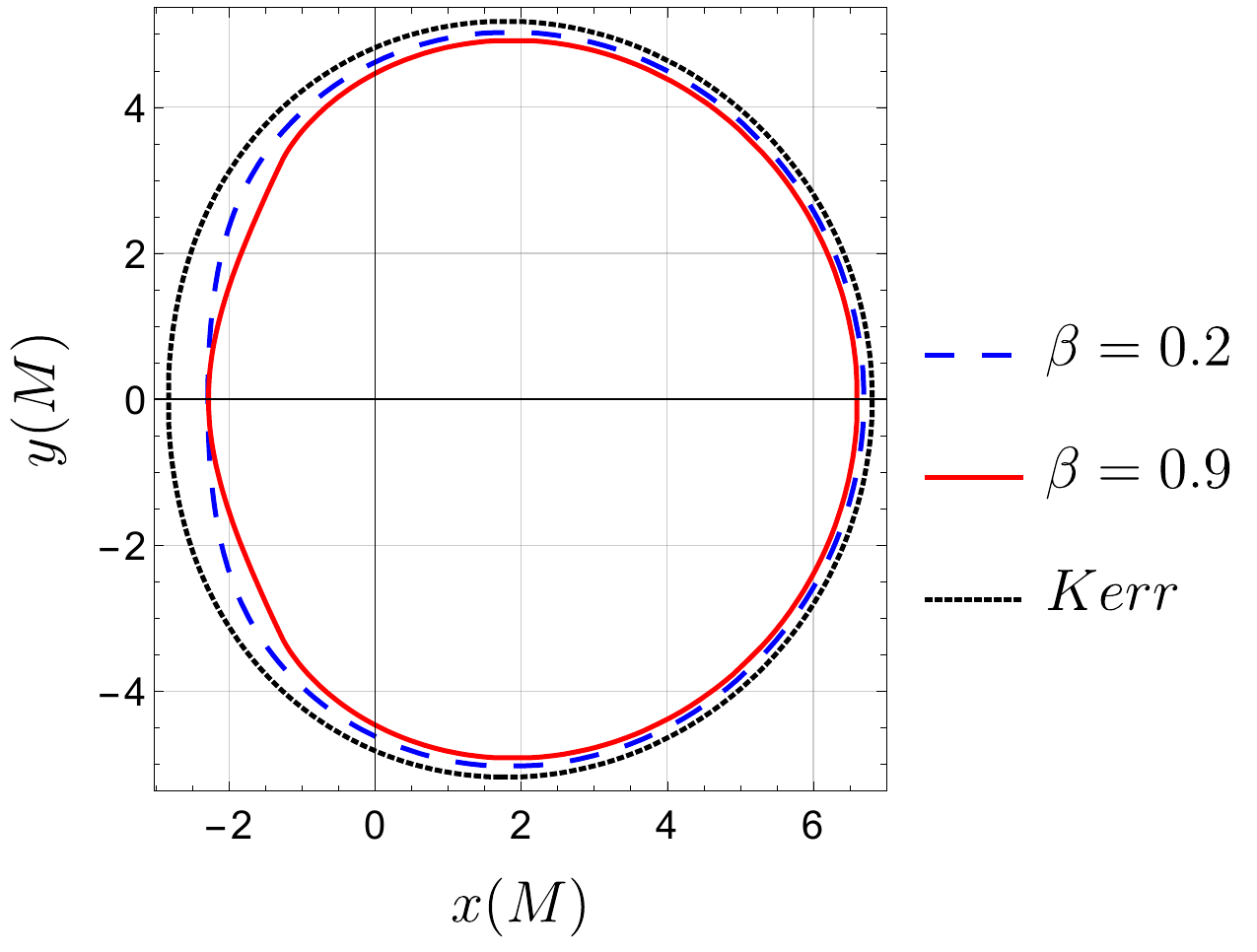}
        \caption{$\theta_0=\pi/2.$}
        \label{fig:}
    \end{subfigure}
\hspace{0.8 mm}
    \begin{subfigure}[b]{0.45\textwidth}
        \includegraphics[width=\textwidth]{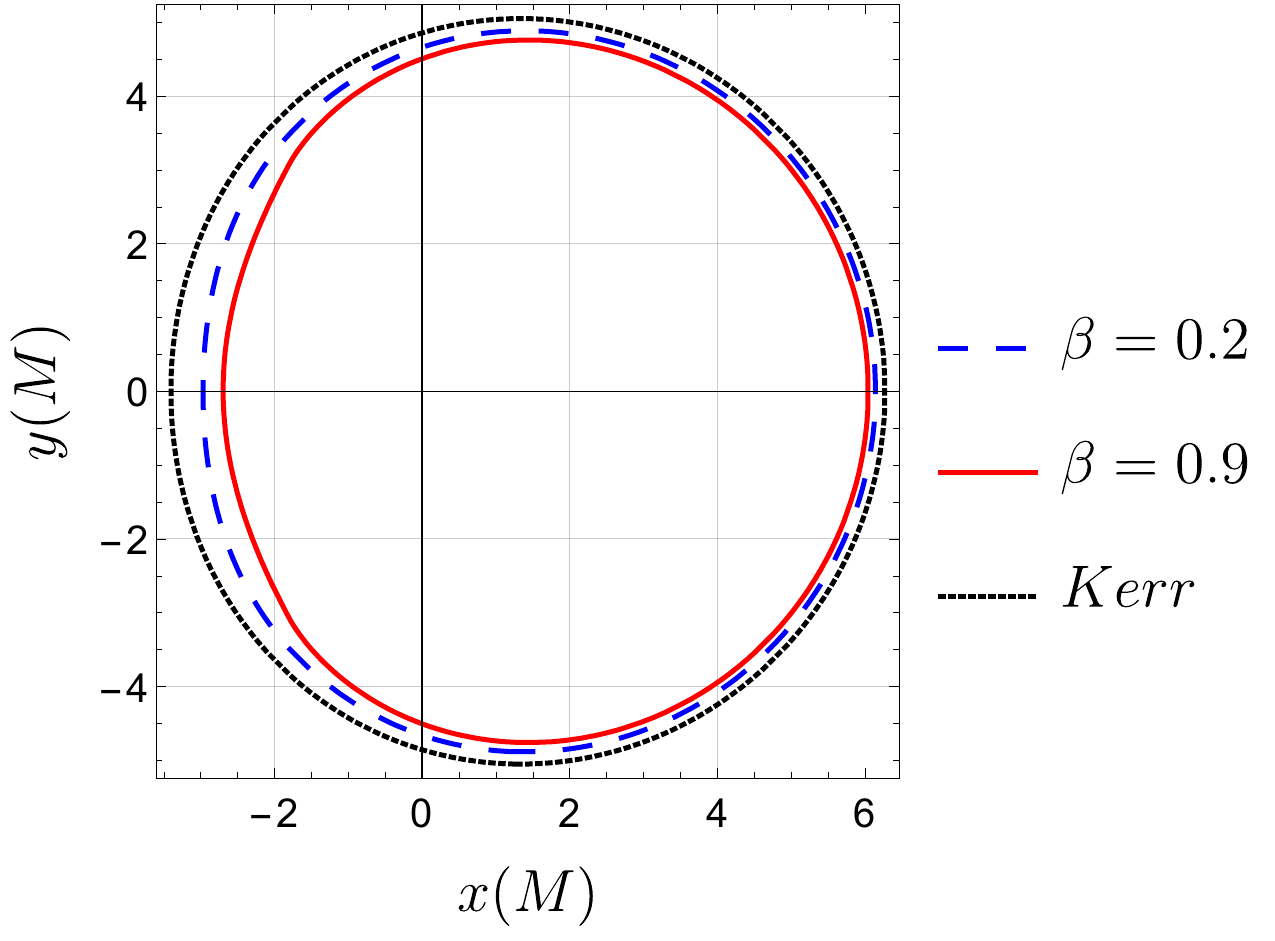}
        \caption{ $\theta_0=\pi/4.$}
        \label{fig:TS_b_small_All_a--0p9}
    \end{subfigure}
    \caption{Black hole shadows in the weak non-minimal coupling regime for $M=1$, $Q=0.4$, $a=0.9$. Here $\theta_0$ is the angle of inclination.}\label{fig:TS_b_small_All_a--0p9}
\end{figure}
\subsection{Shadows for $\beta>1$, $a<M$ (strong non-minimal coupling regime)}

\begin{figure}[H]
    \centering
    \begin{subfigure}[b]{0.45\textwidth}
        \includegraphics[width=\textwidth]{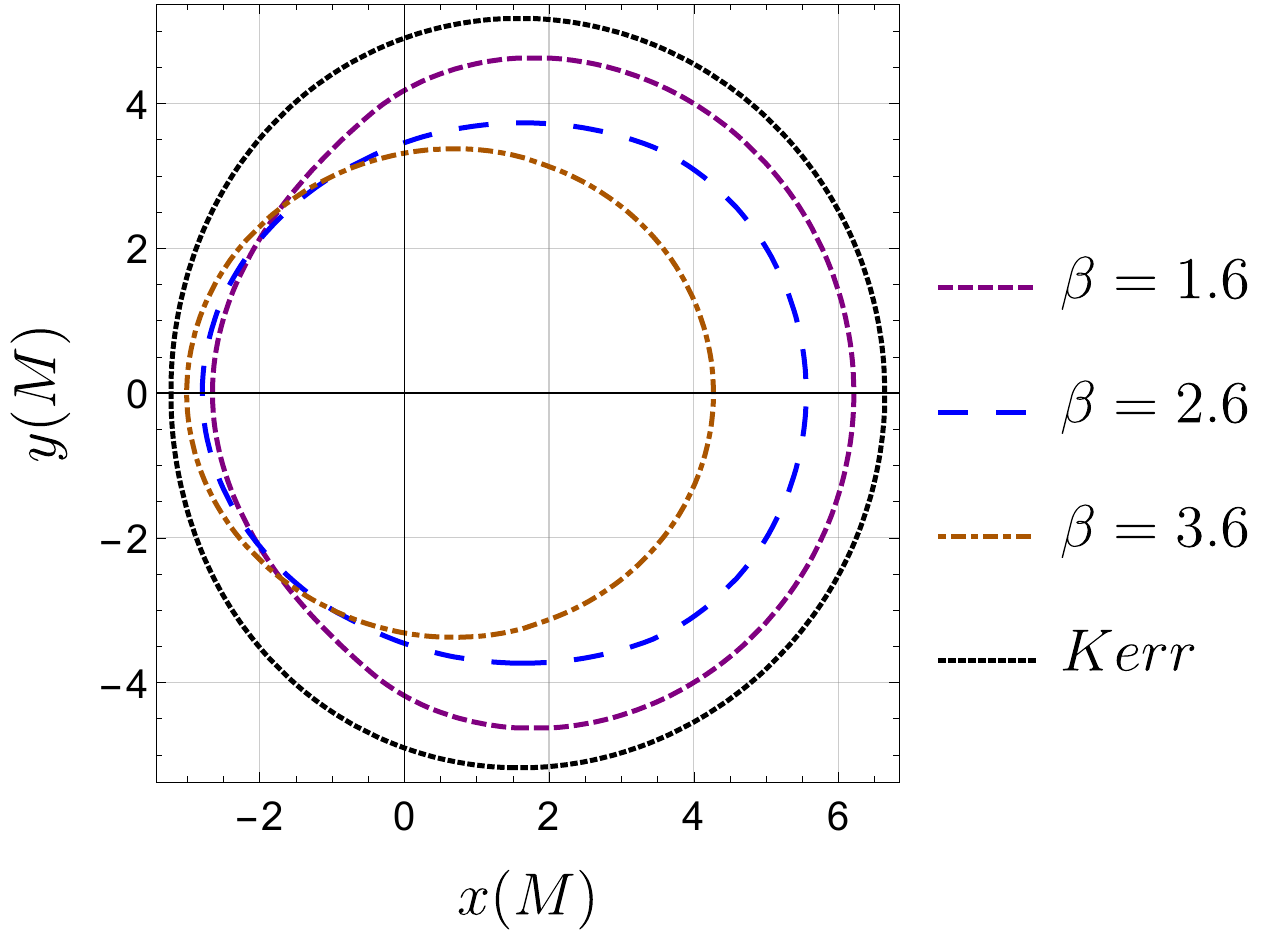}
        \caption{ $\theta_0=\pi/2.$}
        \label{fig:}
    \end{subfigure}
\hspace{0.8 mm}
    \begin{subfigure}[b]{0.45\textwidth}
        \includegraphics[width=\textwidth]{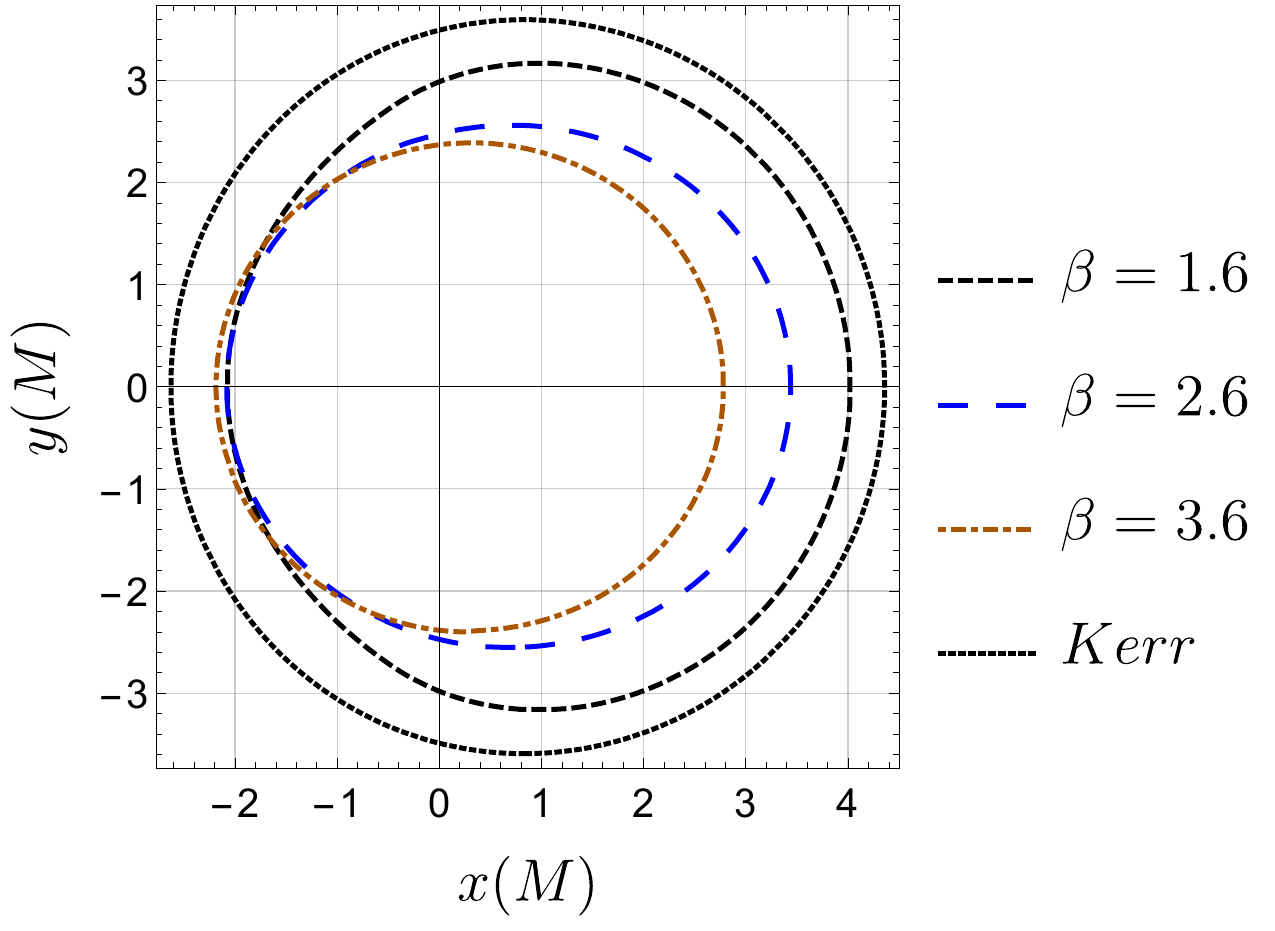}
        \caption{ $\theta_0=\pi/4.$}
        \label{fig:TS_b_All_a--0p8}
    \end{subfigure}
    \caption{Black hole shadows in the strong non-minimal coupling regime for $M=1$, $Q=0.4$, $a=0.8$.}\label{fig:TS_b_All_a--0p8}
\end{figure}
\begin{figure}[H]
    \centering
    \begin{subfigure}[b]{0.45\textwidth}
        \includegraphics[width=\textwidth]{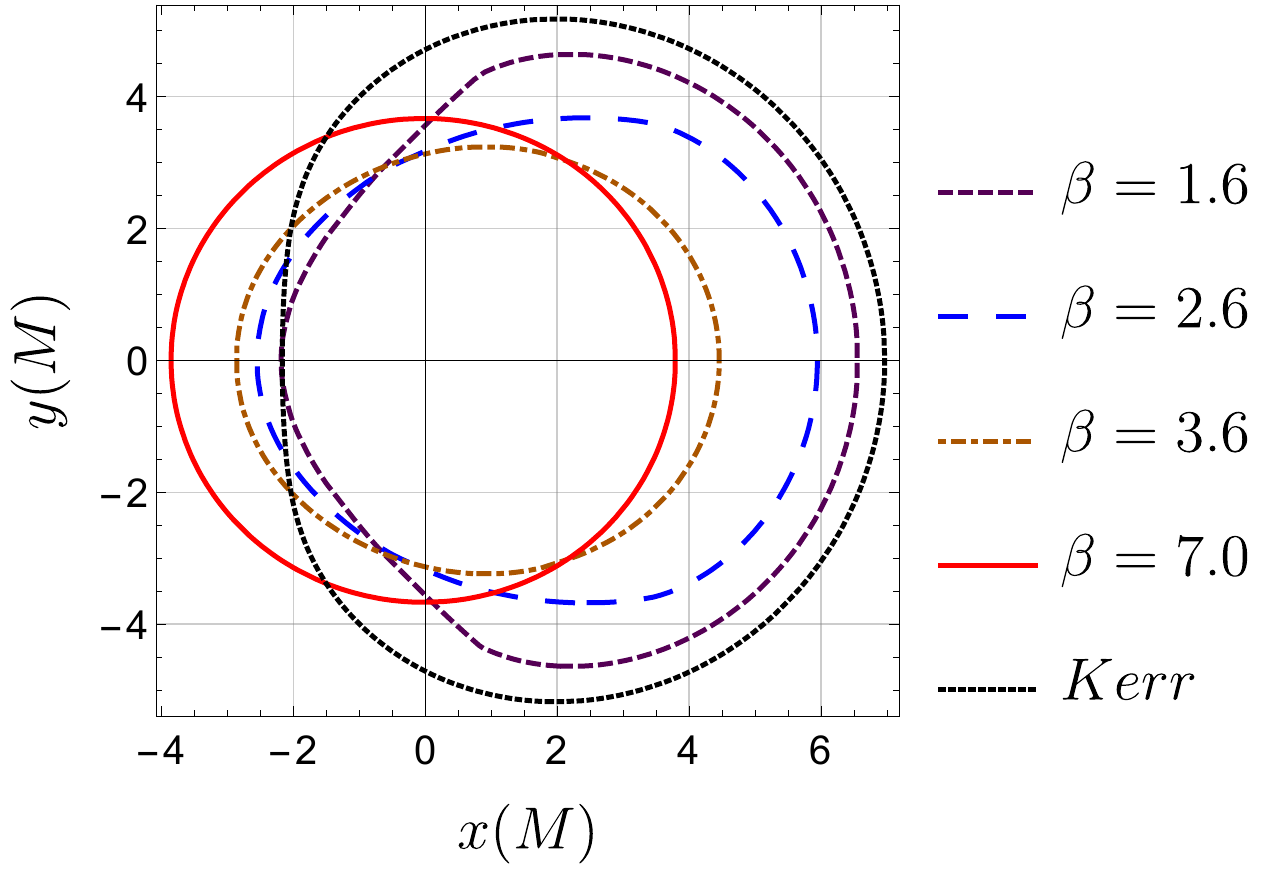}
        \caption{ $\theta_0=\pi/2.$}
        \label{fig:}
    \end{subfigure}
\hspace{0.8 mm}
    \begin{subfigure}[b]{0.45\textwidth}
        \includegraphics[width=\textwidth]{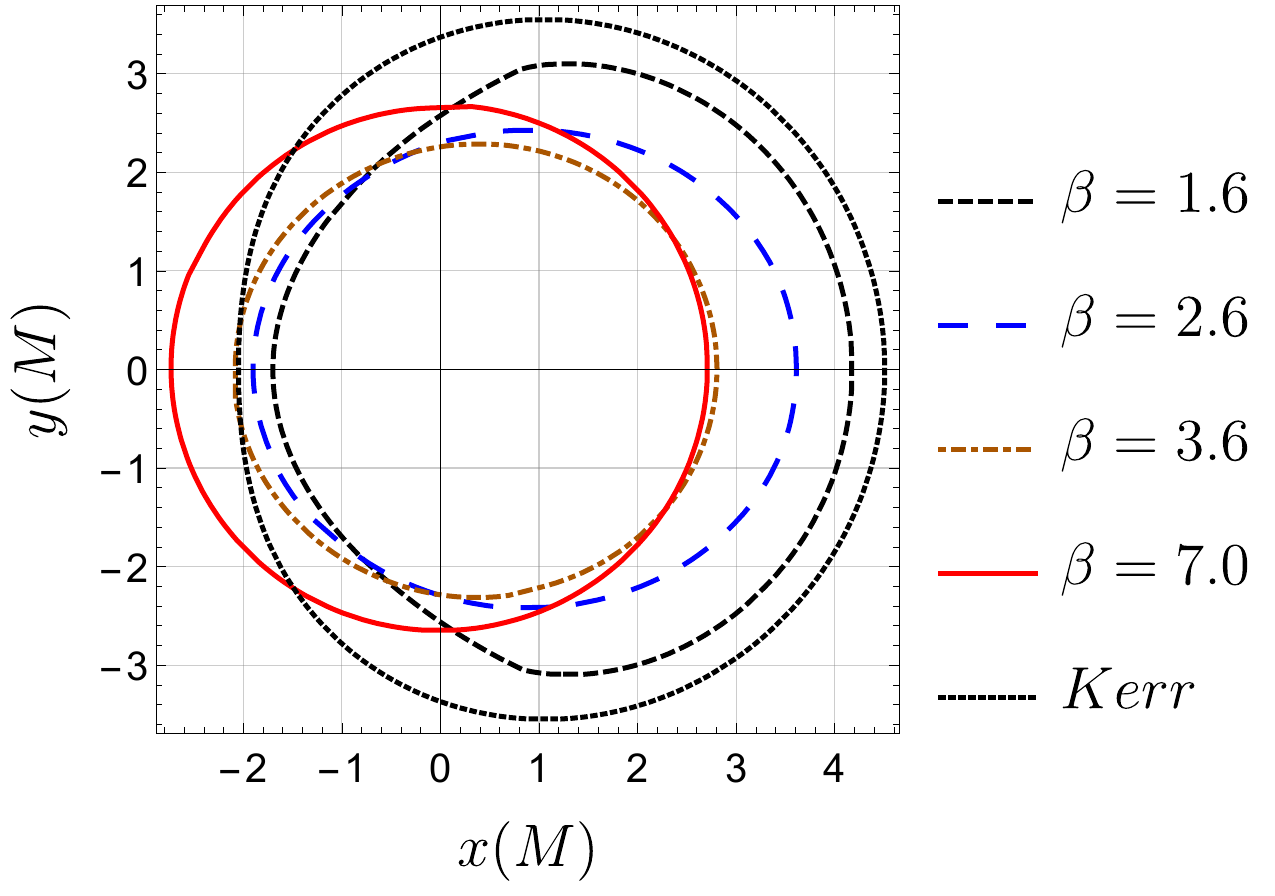}
        \caption{ $\theta_0=\pi/4.$}
        \label{fig:}
    \end{subfigure}
    \caption{Black hole shadows in the strong non-minimal coupling regime for $M=1$, $Q=0.4$, $a=0.995$.}\label{fig:TS_Pi-4_b_All_a--0p995}
\end{figure}

\subsection{Shadows for $\beta>1$, $a>M$ (ultraspinning strong non-minimal coupling regime)}

\begin{figure}[H]
    \centering
    \begin{subfigure}[b]{0.45\textwidth}
        \includegraphics[width=\textwidth]{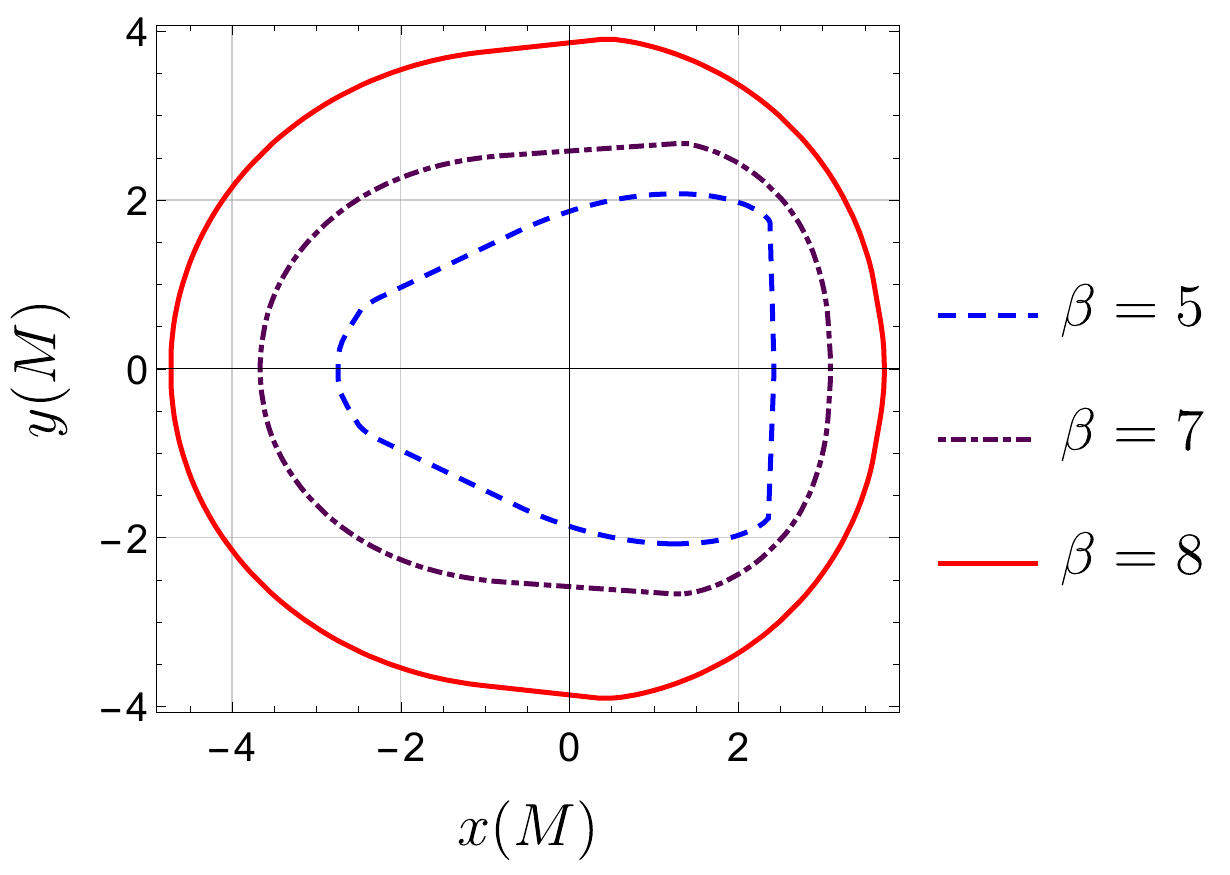}
        \caption{ $\theta_0=\pi/2$.}
        \label{fig:}
    \end{subfigure}
\hspace{0.8 mm}
    \begin{subfigure}[b]{0.45\textwidth}
        \includegraphics[width=\textwidth]{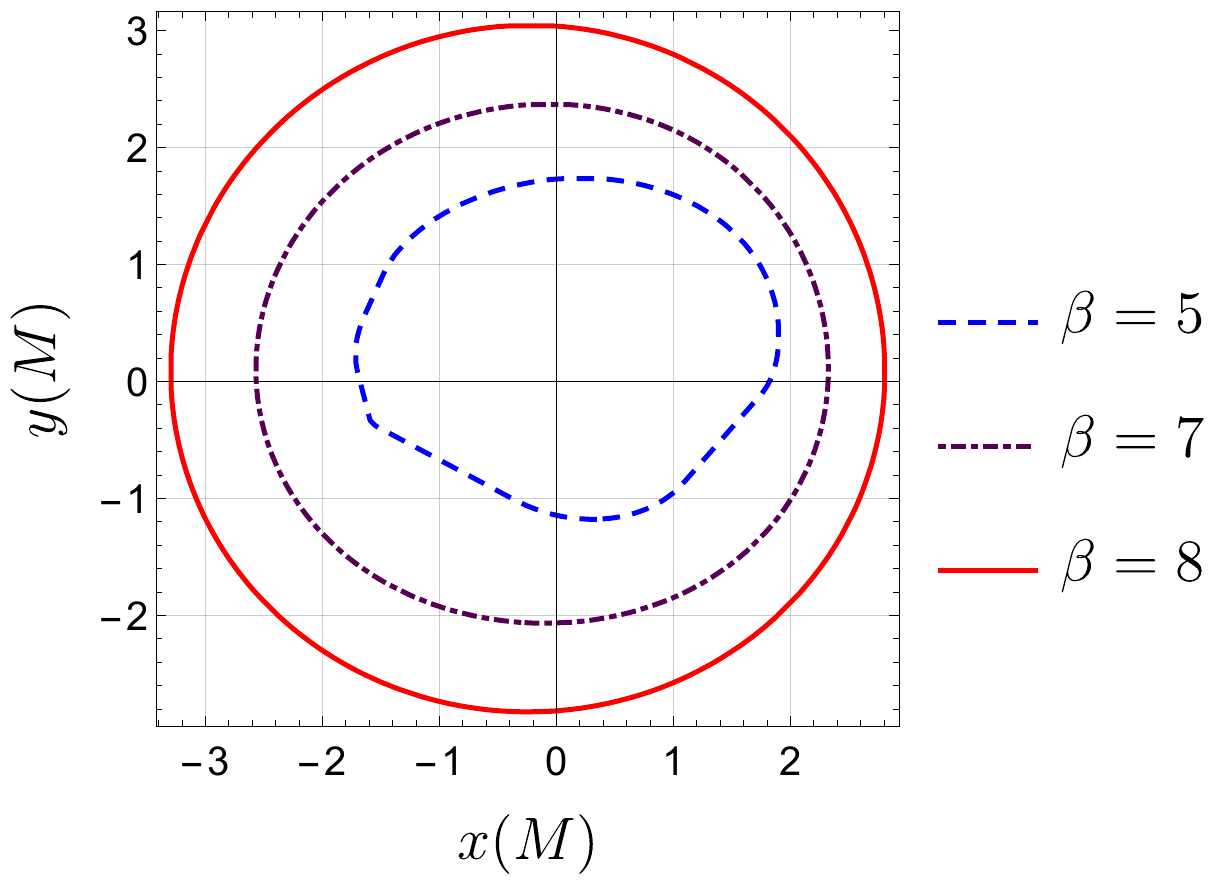}
        \caption{ $\theta_0=\pi/4$.}
        \label{fig:}
    \end{subfigure}
    \caption{Black hole shadows in the  ultraspinning strong non-minimal coupling regime for $M=1$, $Q=0.4$, $a=2.0$.}\label{fig:TS_Pi-4_b_All_a--2p0}
\end{figure}
\begin{figure}[H]
    \centering
    \begin{subfigure}[b]{0.45\textwidth}
        \includegraphics[width=\textwidth]{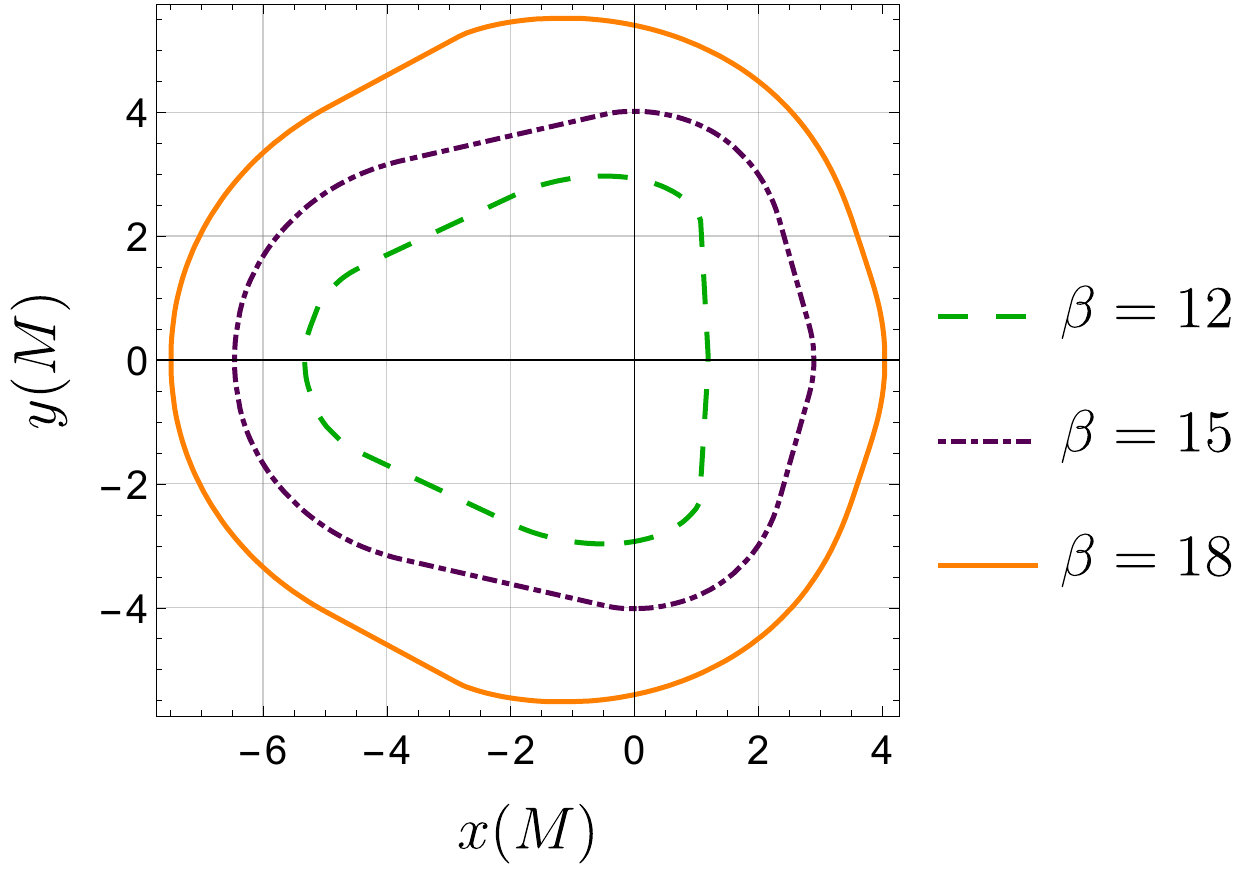}
        \caption{ $\theta_0=\pi/2$.}
        \label{fig:}
    \end{subfigure}
\hspace{0.8 mm}
    \begin{subfigure}[b]{0.45\textwidth}
        \includegraphics[width=\textwidth]{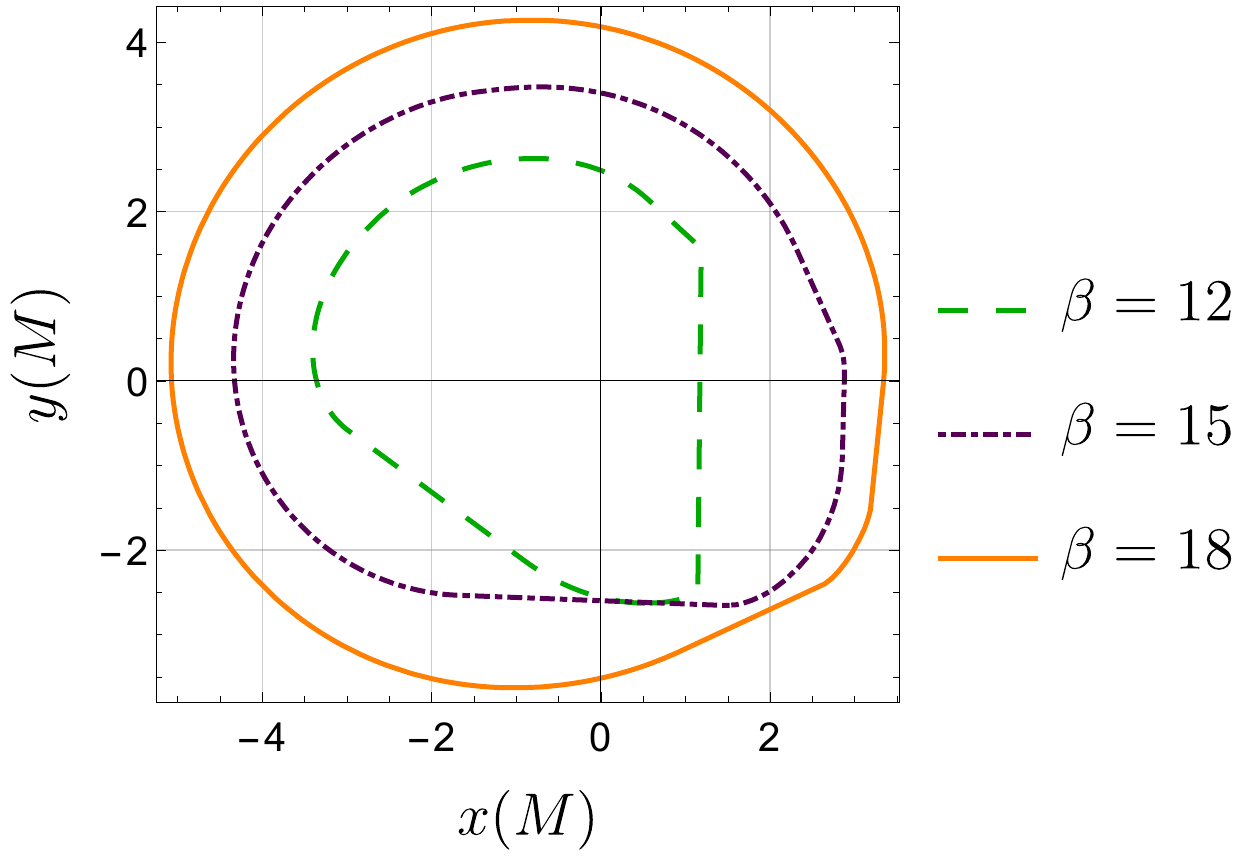}
        \caption{ $\theta_0=\pi/4$.}
        \label{fig:}
    \end{subfigure}
    \caption{Black hole shadows in the  ultraspinning strong non-minimal coupling regime for $M=1$, $Q=0.4$, $a=4.0$.}\label{fig:TS_Pi-4_b_All_a--4p0}
\end{figure}
\begin{figure}[H]
    \centering
    \begin{subfigure}[b]{0.45\textwidth}
        \includegraphics[width=\textwidth]{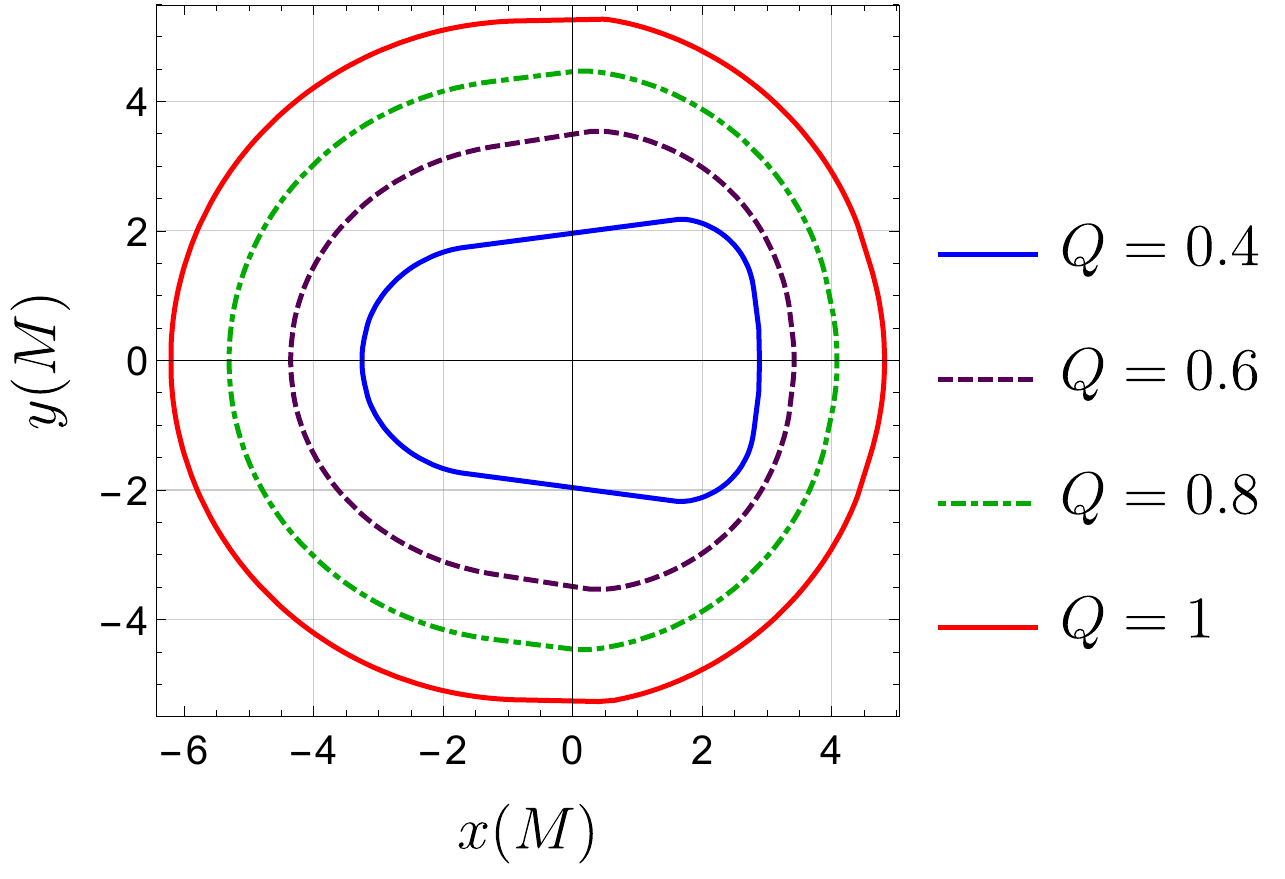}
        \caption{ $\theta_0=\pi/2$.}
        \label{fig:}
    \end{subfigure}
\hspace{0.8 mm}
    \begin{subfigure}[b]{0.45\textwidth}
        \includegraphics[width=\textwidth]{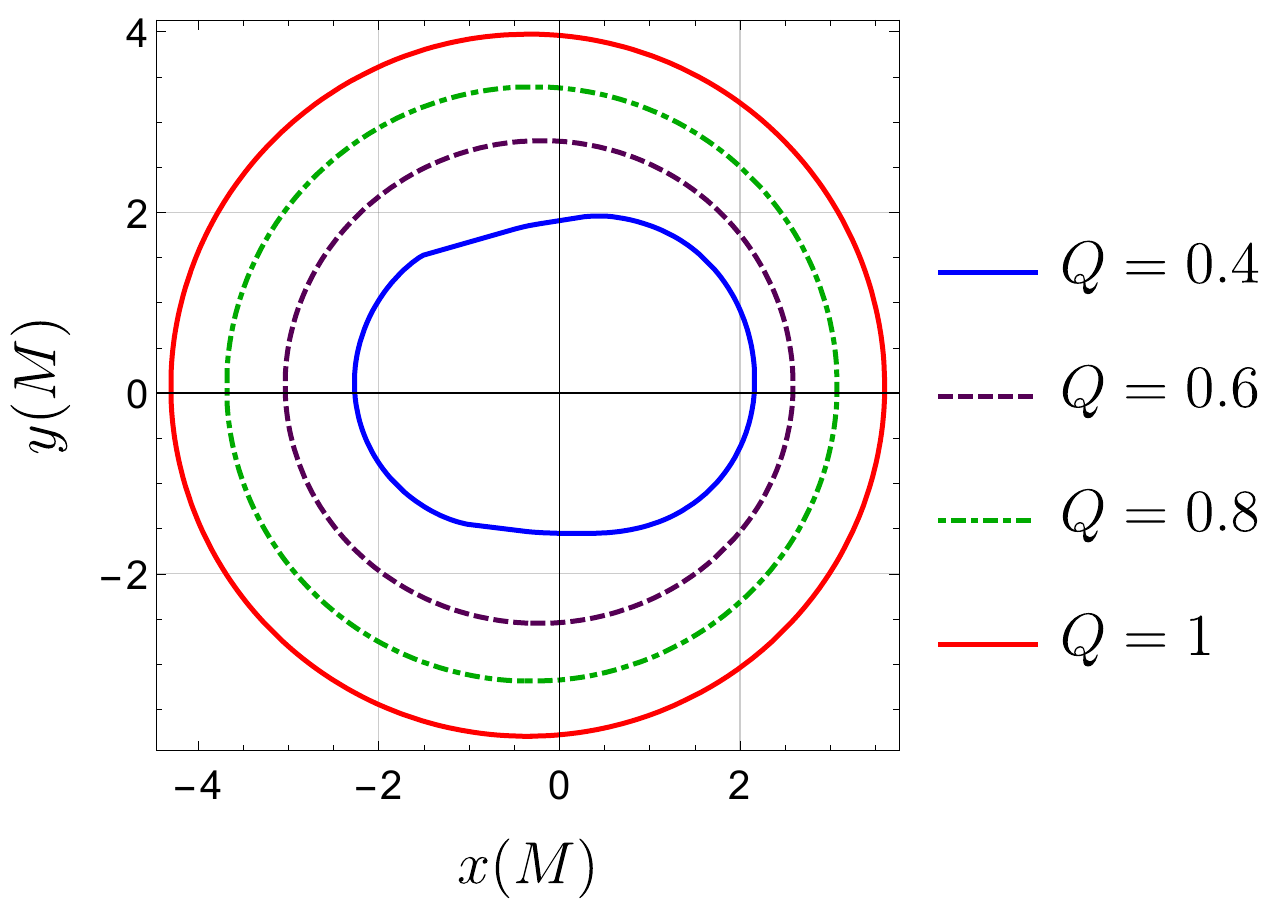}
        \caption{ $\theta_0=\pi/4$.}
        \label{fig:}
    \end{subfigure}
    \caption{Black hole shadows in the  ultraspinning strong non-minimal coupling regime for $M=1$, $\beta=6.0$, $a=2.0$.}\label{fig:TS_Q_All_b--0p6_a--2p0}
\end{figure}

\subsection{Shadows for $\beta>1$, $M=0$  (massless strong non-minimal coupling regime)}

\begin{figure}[H]
    \centering
    \begin{subfigure}[b]{0.45\textwidth}
        \includegraphics[width=\textwidth]{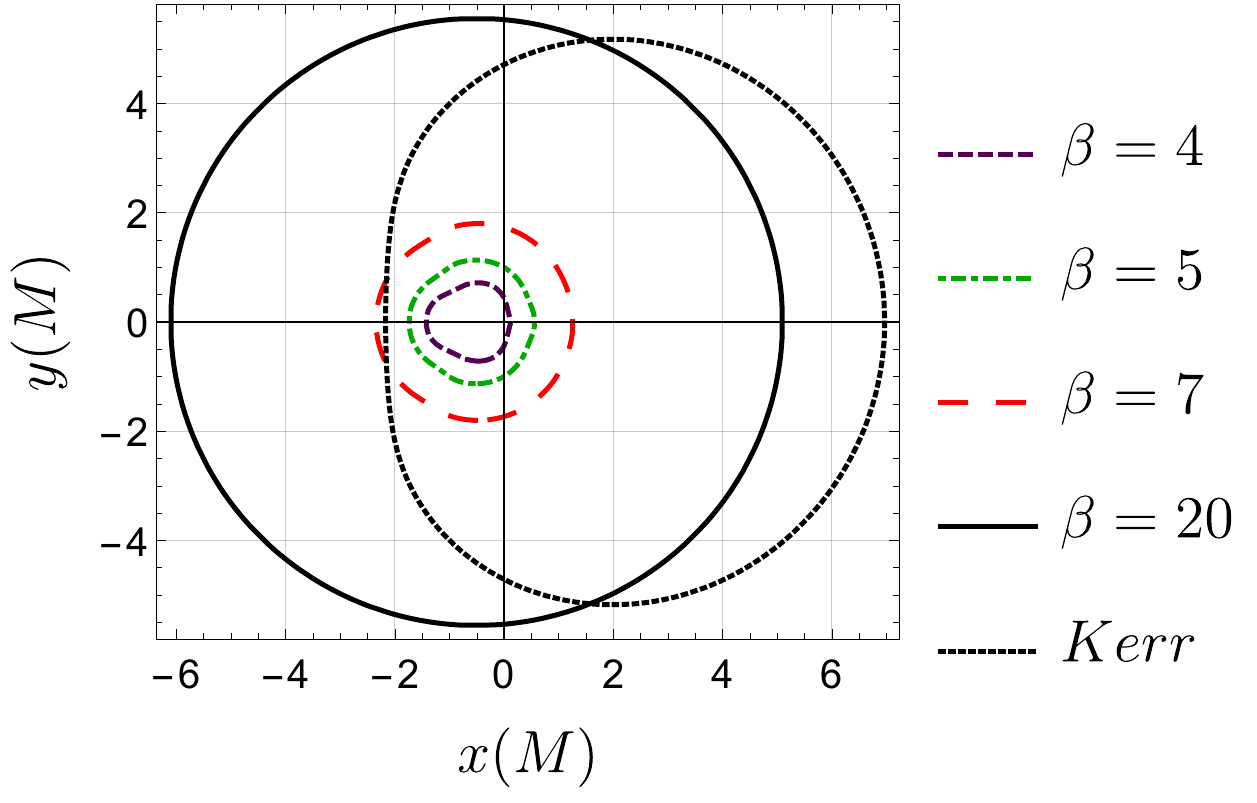}
        \caption{ $\theta_0=\pi/2$.}
        \label{fig:}
    \end{subfigure}
\hspace{0.8 mm}
    \begin{subfigure}[b]{0.45\textwidth}
        \includegraphics[width=\textwidth]{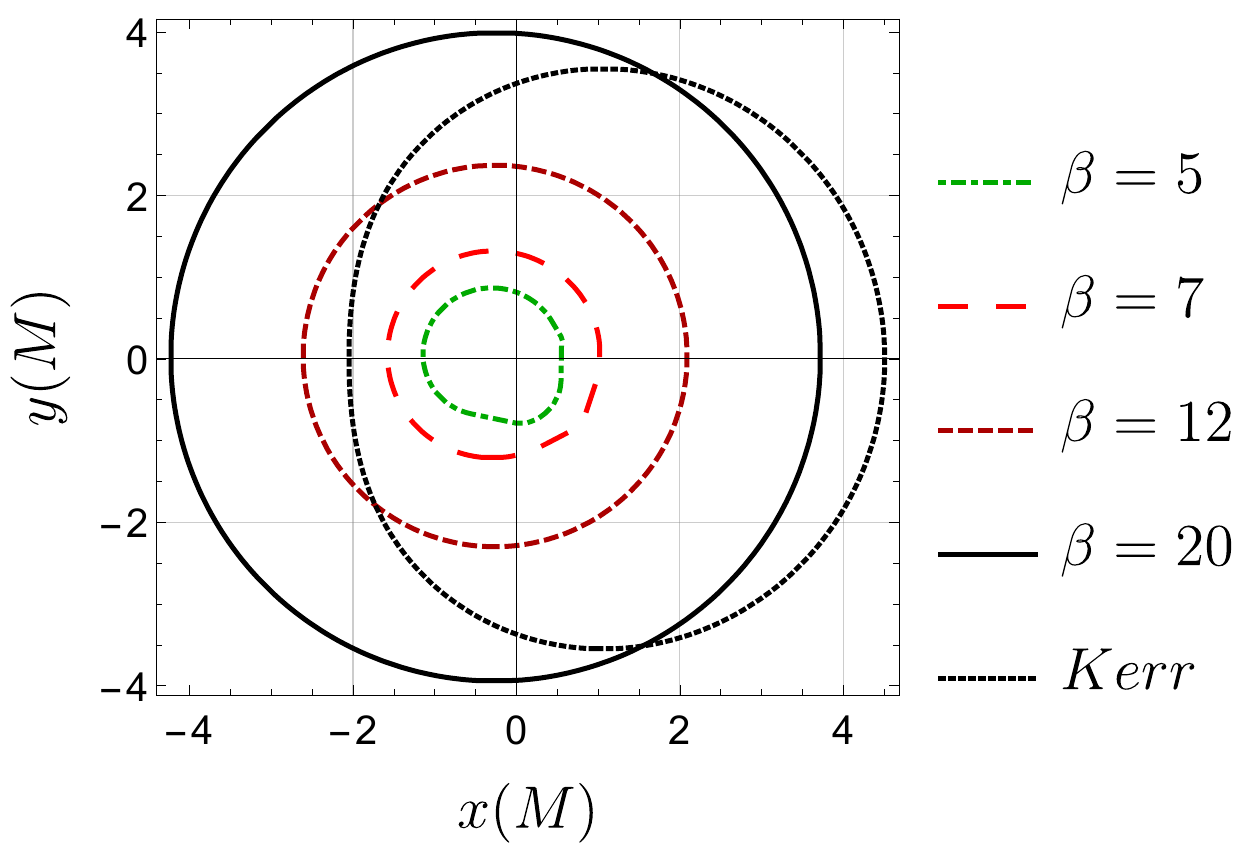}
        \caption{ $\theta_0=\pi/4$.}
        \label{fig:}
    \end{subfigure}
    \caption{Black hole shadows in the  massless strong non-minimal coupling regime for $M=0$, $Q=0.4$, $a=0.995$.}\label{fig:TS_massless_b_All_a--0p995}
\end{figure}

\begin{figure}[H]
    \centering
    \begin{subfigure}[b]{0.45\textwidth}
        \includegraphics[width=\textwidth]{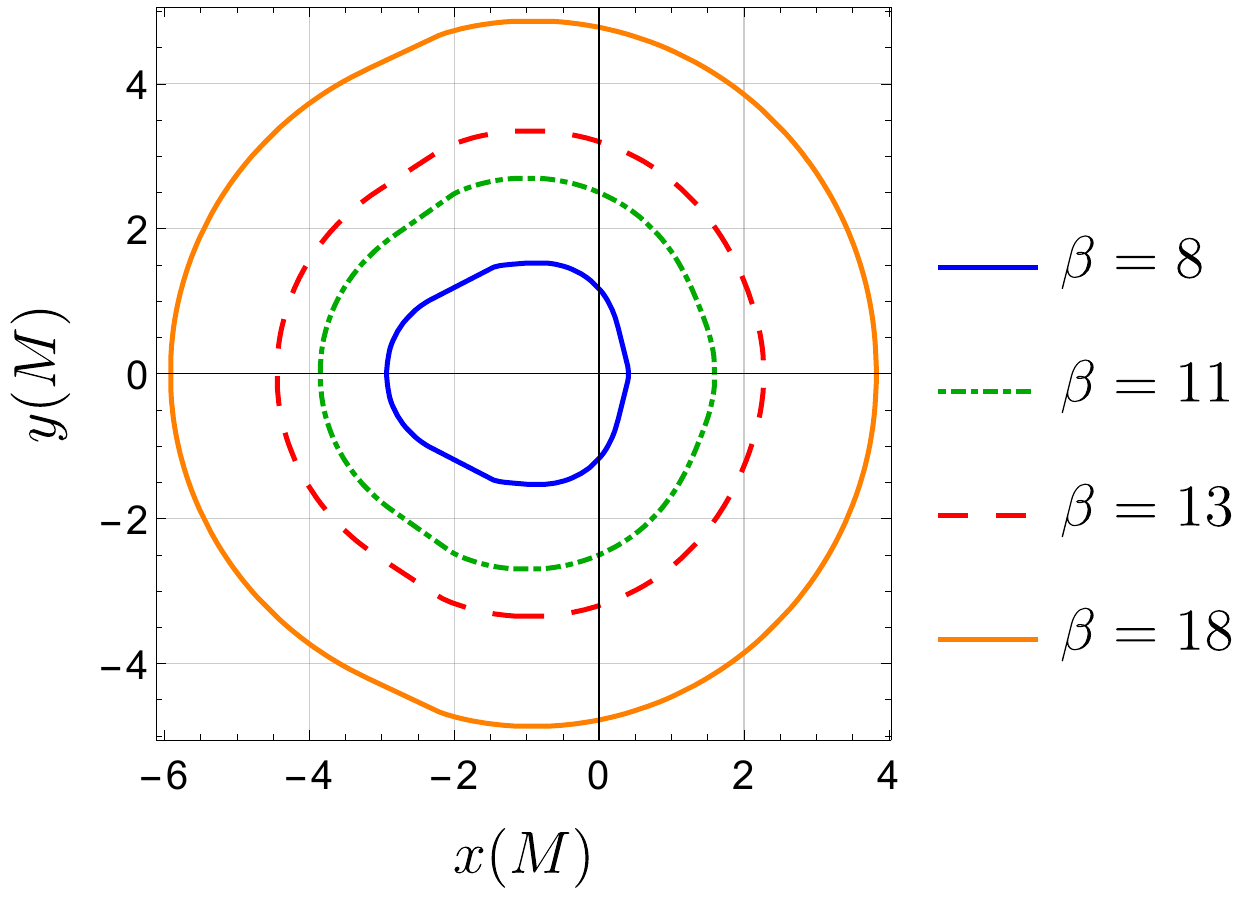}
        \caption{ $\theta_0=\pi/2$.}
        \label{fig:}
    \end{subfigure}
\hspace{0.8 mm}
    \begin{subfigure}[b]{0.45\textwidth}
        \includegraphics[width=\textwidth]{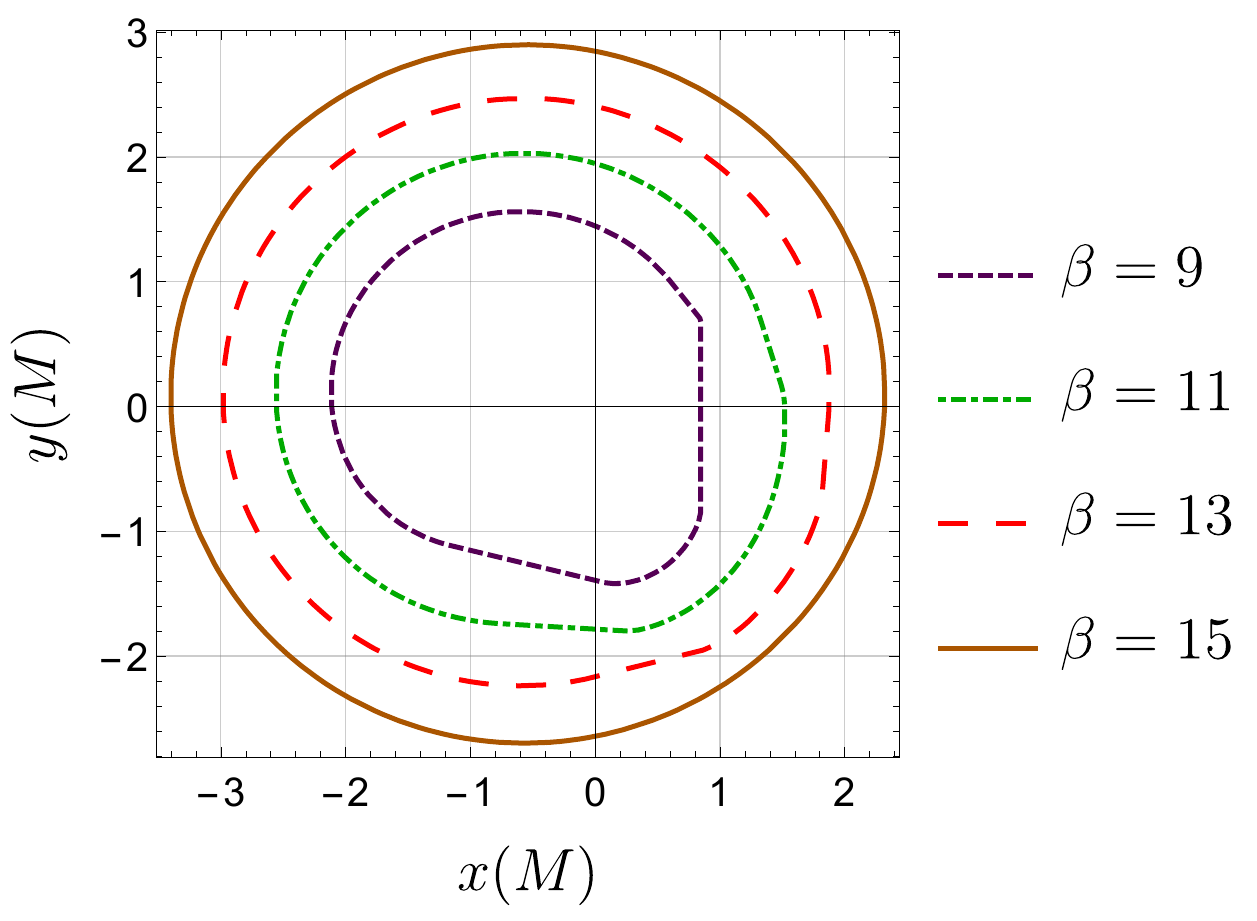}
        \caption{ $\theta_0=\pi/4$.}
        \label{fig:}
    \end{subfigure}
    \caption{Black hole shadows in the  ultraspinning massless strong non-minimal coupling regime for $M=0$, $Q=0.4$, $a=2.0$.}\label{fig:TS_massless_Pi-4_b--All_a--2p0}
\end{figure}

\begin{figure}[H]
    \centering
    \begin{subfigure}[b]{0.45\textwidth}
        \includegraphics[width=\textwidth]{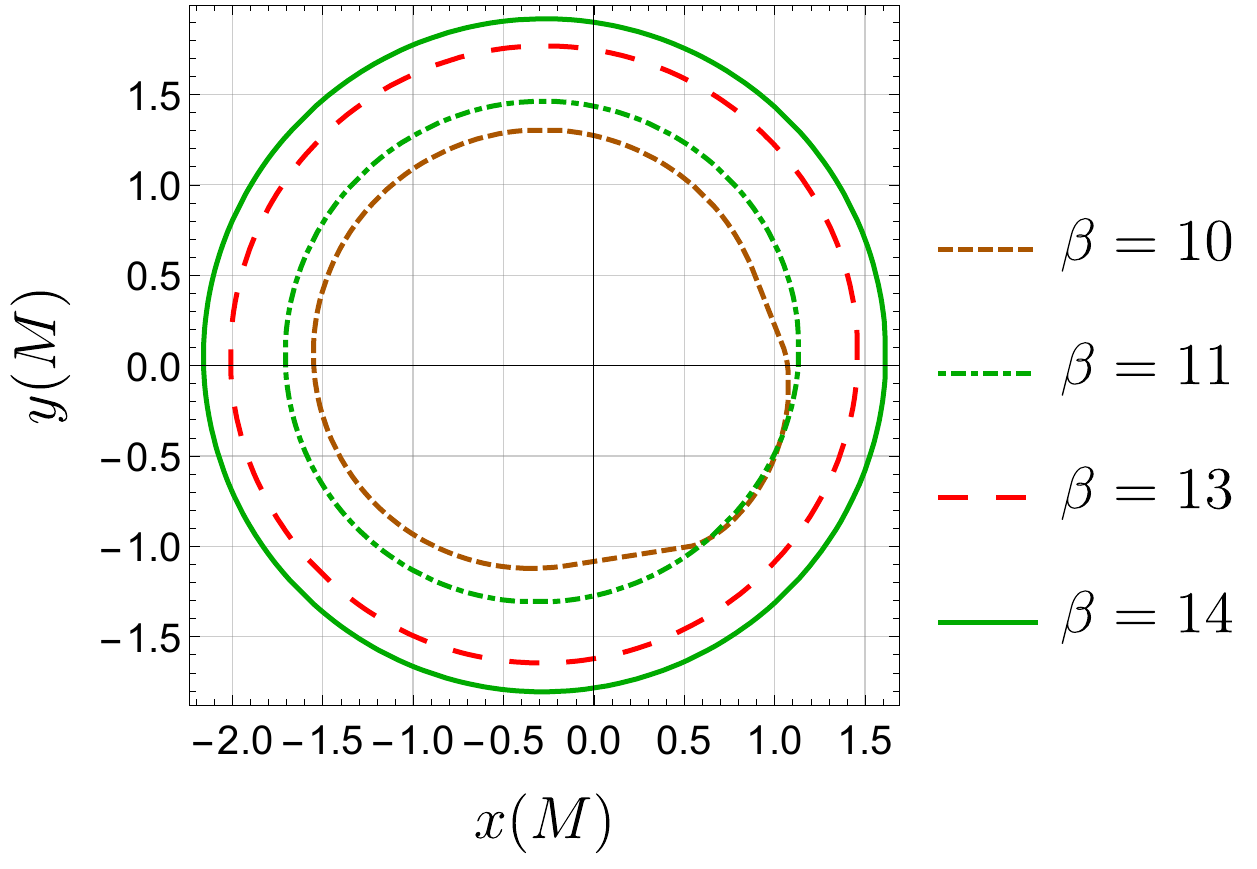}
        \caption{ $\theta_0=\pi/6$.}
        \label{fig:}
    \end{subfigure}
\hspace{0.8 mm}
    \begin{subfigure}[b]{0.45\textwidth}
        \includegraphics[width=\textwidth]{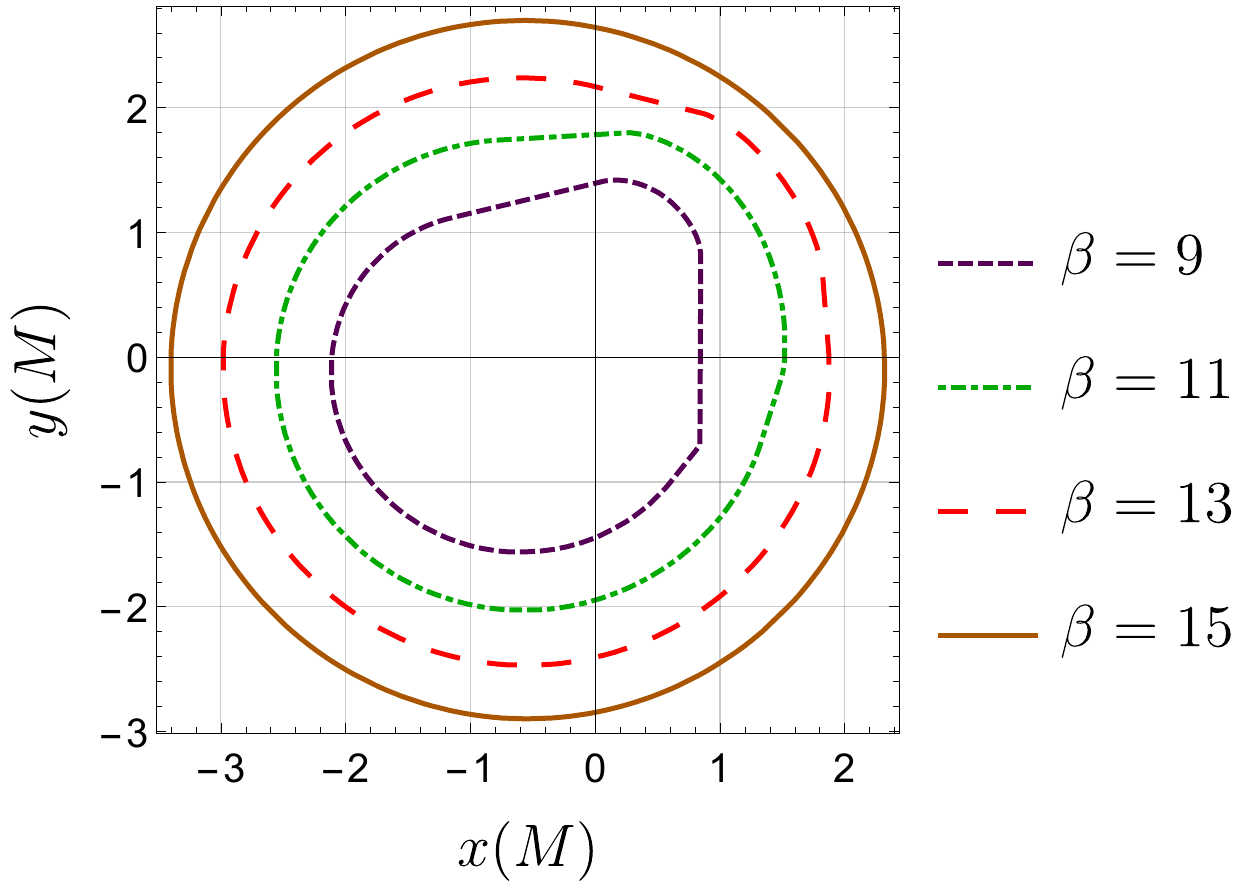}
        \caption{ $\theta_0=3\,\pi/4$.}
        \label{fig:}
    \end{subfigure}
    \caption{Black hole shadows in the  ultraspinning massless strong non-minimal coupling regime for $M=0$, $Q=0.4$, $a=2.0$.}\label{fig:TS_massless_3-Pi-4_b--All_a--2p0}
\end{figure}

\section{Conclusion}\label{sec:conclusion}

The analysis of the horizon as function of the Boyer-Lindquist angle $\theta$, depicted on Fig. \ref{fig:TS_rth1}, shows some interesting features. When considering the case $a<M$, one notices that the outer horizon increases from the north pole to the black hole equator and decreases symmetrically from the equator to the south pole. Meanwhile, the Cauchy horizon has an opposite behaviour.

When considering the ultraspinning case, $a>M$, one notices that for some values of the disformal parameter $\beta$ there are values of $\theta$ for which the horizons don't exist. There are exactly two specific values of $\theta\in[0,\pi]$, one in the northern hemisphere, and one in the southern hemisphere, where the two horizons coincide and an extremal black hole is formed. Each of these characteristics of the both horizons can be clearly seen in Fig. \ref{fig:TS_rth3}, where the three-dimensional shape of the horizons and their polar cross-sections are shown.

On the equatorial plane, $\theta=\pi/2$, the Eq. (\ref{HorizonEq}) reduces to two quadratic equations, namely, one trivial $r^2=0$, and one non-trivial $r^2-2Mr+a^2+Q^2(1-\beta^2)=0$ with roots $r_{\pm}=M\pm\sqrt{M^2-a^2+Q^2(1-\beta^2)}$. In this case the radius of the event horizon is maximal, while the radius of the Cauchy horizon is minimal and is exactly zero when $a^2=Q^2(1-\beta^2)$ or $\beta^2=1-a^2/Q^2$.

Now let us consider the dependence of the horizons on the non-minimal coupling $\beta$ at different angles $\theta$ (Fig. \ref{fig:TS_rth2}). The radius of the event horizon always increases, while the radius of the Cauchy horizon always decreases, with the increasing of $\beta$. In this case there exist one unique value for $\beta$ for which the two horizons coincides. Below this value no black hole exists. On Fig. \ref{fig:TS_rth2} one observes again that the radius of the event horizon monotonically increases from the north pole to the equator. Furthermore, from Table \ref{table:one} and Table \ref{table:two} one notes that the oblateness of the black hole gets smaller when $\beta$ increases until  the outer horizon approaches a sphere for very large $\beta$.

The shadow of the black hole in the weak non-minimal coupling regime (Fig. \ref{fig:TS_b_small_All_a--0p9}) is getting bigger with increasing $\beta$, but always stays within the silhouette of the Kerr black hole shadow. It approaches the Kerr-Newman black hole shadow contour at $\beta=0$ and the Kerr black hole shadow at $\beta=1$. Furthermore, the cuspy silhouette of the black hole image is becoming more apparent when $\beta$ approaches 1.

The shadow of the black hole in the case of $a<M$ in the strong non-minimal coupling regime (Fig. \ref{fig:TS_b_All_a--0p8}) at first decreases with increasing $\beta$, but after some value of $\beta$ the size of the shadow begins to increase (Fig. \ref{fig:TS_Pi-4_b_All_a--0p995}), which can also be depicted from the values of the mean radius $\bar r_{sh}$ of the shadow and its perimeter $\mathcal{P}$,  given in Table \ref{table:one}. This is an interesting phenomenon, because it was unexpected and we did not observe it in the weak non-minimal coupling regime or the ultraspinning regime, where the shadow always increases with increasing values of $\beta$. A possible explanation is due to the non-minimal coupling of the dark vector field to gravity, the underlying mechanisms for which are unclear at the moment. This phenomenon poses questions that can be answered by studying the gravitational lensing effect in the strong deflection limit nearby the black hole photon orbits. The expected results of such a survey can provide valuable information about the space-time type around the compact object and will be reported soon in a future work. Finally, one notices that the cuspy silhouette of the black hole shadow vanishes for $\beta \gg 1$ and also for smaller values of the black hole angular momentum, which is supported by the decreasing values of $\sigma_r$ given in Table \ref{table:one}.

In the ultraspinning case (Fig. \ref{fig:TS_Pi-4_b_All_a--2p0} and Fig. \ref{fig:TS_Pi-4_b_All_a--4p0}) the apparent image of the black hole in the observer's sky is highly deformed for all values of the angle of inclination $\theta_{0}$. The size of the shadow increases with increasing $\beta$, but the cusps remain yet visible. Furthermore, for even larger spin (Fig. \ref{fig:TS_Pi-4_b_All_a--4p0}) the shadow gets asymmetrically deformed with respect to the horizontal abscissa of the observer's plane.

In the massless case (Figs. \ref{fig:TS_massless_b_All_a--0p995}-\ref{fig:TS_massless_3-Pi-4_b--All_a--2p0}) the size of the shadow starts from very small (within the Kerr shadow) when $\beta<1$ and increases beyond the size of the Kerr black hole shadow for $\beta\gg 1$. Also the shape of the shadow approaches spherical form when $\beta\gg 1$, which is confirmed by the values of $\sigma_r$ parameter in Table \ref{table:two}, valid for the massless as well as the massive case.

Finally, the increasing values of the charge $Q$, at fixed $\beta$, lead to smoother and larger silhouettes of the black hole shadow, which are portrayed in Fig. \ref{fig:TS_Q_All_b--0p6_a--2p0}.

In all considered cases one notices that the center of the black hole $x_C$ moves to the left with respect to the center of the Kerr black hole for increasing values of $\beta$. This is confirmed by the data given in Tables \ref{table:one} and \ref{table:two}. Furthermore, the non-equatorial observer, $\theta_{0}\neq\pi/2$, will see the same shape of the black hole shadow, no matter it is positioned below or above the equatorial plane, as long as it fulfils the observer condition $\theta_{0}^{\rm north}+\theta_{0}^{\rm south}=\pi$. An examples are shown in Fig. \ref{fig:TS_massless_Pi-4_b--All_a--2p0} (b) and Fig. \ref{fig:TS_massless_3-Pi-4_b--All_a--2p0} (b).
\begin{table}[H]
\centering
  \begin{tabular}{||c c c c c c c c c c c||}
 \hline
 $\beta$ & $\bar r_{sh}$ & $\delta_{\bar r_{sh}}(\%)$ & $r^{pol}_h$ &  $r^{eq}_h$ & $\omega (\%)$ & $\sigma_r$ & $\delta_{\sigma_r}(\%)$ & $\mathcal{P}$ & $x_C$ & $\delta_{x_C} (\%)$ \\ [0.5ex]
 \hline\hline
 Kerr & 7.2 & 0 & 1.10 & 1.10 & 0 & 0.213 & 0 & 31.0 & 2.4 & 0\\
 1.6 & 6.8 & -5.3 & 1.34 & 1.51 & 11.4 & 0.234 & 9.7 & 27.8 & 2.2 & -8  \\
 2.6 & 5.9 & -17.8 & 1.83 & 1.97 & 7.0 & 0.219 & 2.8 & 25.0 & 1.75 & -27  \\
 3.6 & 4.9 & -31.4 & 2.26 & 2.39 & 5.3 & 0.125 & -41.1 & 21.7 & 0.85 & -65 \\
 7.0 & 5.2 & -27.2 & 3.67 & 3.77 & 2.6 & 0.019 & -91.2 & 23.4 & 0 & -100 \\
 10.0 & 6.2 & -13.5 & 4.90 & 4.98 & 1.6 & 0.024 & -88.6 & 27.9 & -0.2 & -108 \\ [1ex]
 \hline
 \end{tabular}
 \caption{Black hole's polar and equatorial radial sizes of the event horizon $r_{h}^{pol}$ and $r_{h}^{eq}$, the oblatness $\omega$, as well as the shadow's mean radius $\bar r_{sh}$, perimeter $\mathcal{P}$, the black hole center abscissa $x_C$ and equatorial relative shadow deviations from Kerr  for fixed parameters: $M=1$, $\theta_0=\pi/2$, $a=0.995$. }
 \label{table:one}
\end{table}
\begin{table}[H]
\centering
  \begin{tabular}{||c c c c c c c c||}
 \hline
 $\beta$ & $\bar r_{sh}$ & $r^{pol}_h$ &  $r^{eq}_h$ & $\omega (\%)$ & $\sigma_r$ & $\mathcal{P}$ & $x_C$  \\ [0.5ex]
 \hline\hline
 8.0 & 4.9 & 2.97 & 3.66 & 18.8 & 0.076 & 21.6 & -0.40 \\
 10.0 & 5.7 & 4.13 & 4.58 & 9.9 & 0.065 & 25.7 & -0.50  \\
 12.0 & 6.5 & 5.10 & 5.46 & 6.6 & 0.064 & 29.3 & -0.60  \\
 15.0 & 7.8 & 6.45 & 6.73 & 4.2 & 0.059 & 34.8 & -0.65  \\
 18.0 & 9.0 & 7.74 & 7.98 & 3.0 & 0.056 & 40.1 & -0.75 \\ [1ex]
 \hline
 \end{tabular}
 \caption{Black hole's polar and equatorial radial sizes of the event horizon $r_{h}^{pol}$ and $r_{h}^{eq}$, the oblatness $\omega$, as well as the shadow's mean radius $\bar r_{sh}$, its  perimeter $\mathcal{P}$ and the black hole center abscissa $x_C$ for fixed parameters: $M=1$, $\theta_0=\pi/2$, $a=2.0$. }
 \label{table:two}
\end{table}

\section*{Acknowledgements}

The authors would like to thank D. Doneva, P. Nedkova, K. Staykov, B. Lazov and S. Mladenov for their insightful comments. This work was partially supported by the Bulgarian NSF grant \textnumero~DM18/1 and Sofia University Research Fund under Grant \textnumero~3258/2017. The support
by the COST Actions CA15117 and CA16104 is also gratefully acknowledged.







\bibliographystyle{utphys}
\bibliography{TS-refs}

\end{document}